%% file: manuscript.tex
\documentclass[authorproof,JACIII,researchpaper]{fujipressarticle}
\usepackage{graphicx}
\usepackage{cite}

\usepackage{algorithm}
\usepackage{algpseudocode}

\SetVolume{00}
\SetIssueNo{0}
\SetPubYear{0000}

\begin{document}

\pagestyle{fujipressstyle}

\title{Multi-team Formation System for \\ Collaborative Crowdsourcing}
\author{Ryota Yamamoto and Kazushi Okamoto}
\address{Department of Informatics, Graduate School of Informatics and Engineering, The University of Electro-Communications \\
	1-5-1 Chofugaoka, Chofu, Tokyo, Japan \\
	E-mail: kazushi@uec.ac.jp}
\markboth{R. Yamamoto and K. Okamoto}{Multi-team Formation System for Collaborative Crowdsourcing}
\dates{2000/00/00}{2000/00/00}
\maketitle

\begin{abstract}
\noindent For complex crowdsourcing tasks that require collaboration between multiple individuals, teams should be formed by considering both worker compatibility and expertise.
Furthermore, the nature of crowdsourcing dictates the budget for tasks and workers' remuneration, and excessively large team sizes may reduce collaborative performance.
To address these challenges, we propose a heuristic optimization algorithm that leverages social network information to simultaneously form teams with optimized worker compatibility for multiple tasks.
In our approach, historical collaboration is represented as a social network, where the edge weights correspond to explicit ratings of worker compatibility.
In a simulation experiment using synthetic data, we applied Gaussian process regression to examine the relationship between eight experimental parameters and evaluation values, thereby analyzing the output of the proposed algorithm.
To generate the necessary data for regression, we ran the proposed algorithm with experimental parameters that were sequentially estimated using Bayesian optimization.
Our experiments revealed that the evaluation values were extremely low when the team size limit, the degree mean of the social network, and the task budget were set to low values.
The results also indicate that the proposed algorithm outperformed the hill-climbing method under almost all experimental conditions.
In addition, the highest evaluation values were achieved when the simulated annealing temperature decrease rate was approximately 0.9, while smoothing the objective function proved ineffective.
\end{abstract}


\input{tex/section1.tex}
\input{tex/section2.tex}
\input{tex/section3.tex}
\input{tex/section4.tex}
\input{tex/section5.tex}
\input{tex/section6.tex}

\acknowledgments
This work was supported by JSPS KAKENHI Grant Numbers JP21H03553, JP23K26329.

\bibliographystyle{fujipressbib}
\bibliography{reference.bib}



\begin{profile}
  \Name{Ryota Yamamoto}
  \Affiliation{Department of Informatics, Graduate School of Informatics and Engineering, The University of Electro-Communications}
  \Address{1-5-1 Chofugaoka, Chofu, Tokyo 182-8585, Japan}
  \History{
    2021 Received B.E. degree from The University of Electro-Communications \\
    2023 Received M.E. degree from The University of Electro-Communications
  }
  \Works{
    $\bullet$ R. Yamamoto and K. Okamoto, ``Worker Organization System for Collaborative Crowdsourcing,'' Intelligent and Transformative Production in Pandemic Times (Part of Lecture Notes in Production Engineering), pp.13--22, 2023. \\
    $\bullet$ R. Yamamoto and K. Okamoto, ``An Algorithm for Solving Constraint Satisfaction Problems in Concurrent Formation of Expert Teams,'' IEICE Transactions on Information and Systems, Vol.J107-D, No.3, pp.87--97, 2024.
  }
  \Photo{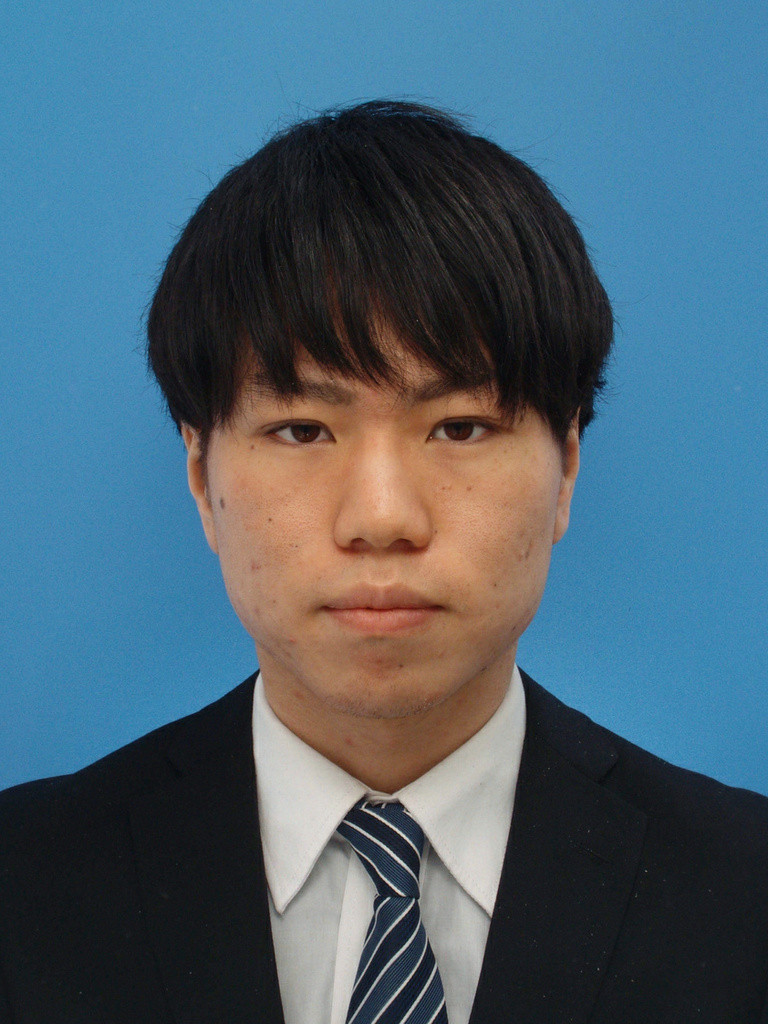}
\end{profile}

\begin{profile}
  \Name{Kazushi Okamoto}
  \Orcid{0000-0002-9571-8909}
  \Affiliation{Department of Informatics, Graduate School of Informatics and Engineering, The University of Electro-Communications}
  \Address{1-5-1 Chofugaoka, Chofu, Tokyo 182-8585, Japan}
  \History{
    2006 Received B.E. degree from Kochi University of Technology \\
    2008 Received M.E. degree from Kochi University of Technology \\
    2011 Received Dr.Eng. degree from Tokyo Institute of Technology \\
    2011-2015 Assistant Professor, Chiba University \\
    2015-2020 Assistant Professor, The University of Electro-Communications \\
    2020- Associate Professor, The University of Electro-Communications
  }
  \Works{
    $\bullet$ K. Sugahara and K. Okamoto, ``Hierarchical Co-clustering with Augmented Matrices from External Domains,'' Pattern Recognition, Vol.142, pp.109657, 2023. \\
    $\bullet$ K. Sugahara and K. Okamoto: ``Hierarchical Matrix Factorization for Interpretable Collaborative Filtering,'' Pattern Recognition Letters, Vol.180, pp.99--106, 2024.
  }
  \Membership{
    $\bullet$ Institute of Electrical and Electronics Engineers (IEEE) \\
    $\bullet$ Association for Computing Machinery (ACM) \\
    $\bullet$ Japan Society for Fuzzy Theory and Systems (SOFT) \\
    $\bullet$ The Institute of Electronics, Information and Communication Engineers (IEICE) \\
    $\bullet$ The Japanese Society for Artificial Intelligence (JSAI)
  }
  \Photo{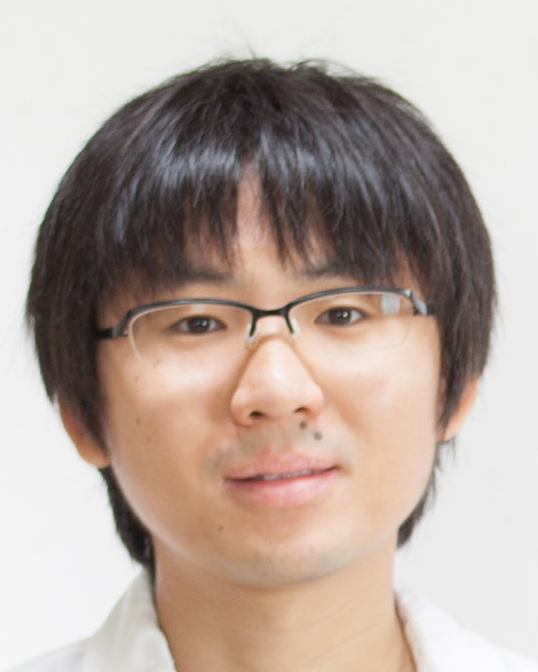}
\end{profile}

\end{document}

%% file: tex/section1.tex
\section{Introduction}
Crowdsourcing is a model in which individuals or organizations delegate tasks to a diverse group of workers recruited via digital platforms.
In this model, it is essential to manage workers with diverse expertise, depending on the scale and nature of the tasks~\cite{2016_h.kashima}.
In particular, collaborative crowdsourcing, in which skilled workers form teams to tackle complex tasks such as software development and translation, has garnered significant attention.
Collaborative teams are expected to have synergistic effects, such as increasing work motivation and complementing each other's expertise~\cite{2011_g.hertel,2011_j.huffmeier}.

To form efficient collaborative teams in crowdsourcing, compatibility is key to ensure easy communication.
Crowd worker teams often consist of individuals from various cultures, workplaces, and time zones, making it challenging to foster belonging and commitment, which can reduce productivity~\cite{2016_l.lykourentzou}.
In team formation, a lack of team communication generally leads to project failure~\cite{2000_k.l.smart}, and team members who have prior experience working with many co-workers are often more communicative~\cite{2012_a.majumder}.

In this study, we address the problem of team formation with optimized team compatibility for crowdsourcing.
Team compatibility refers to the strength of social network connections within the team and is quantified using a social network representing past collaborative experiences.
The nodes in the graph represent workers, and the edges indicate whether past collaboration occurred.
When forming a team for a complex task in crowdsourcing, it is important to consider the following factors:
\begin{itemize}
  \item{Skill levels: Skill-level requirements should be defined. The level required for each task depends on the type of skill, and each worker has different levels of expertise.}
  \item{Hiring costs and budget: The hiring cost of each worker and the budget for each task should be considered to form teams in a crowdsourcing system.}
  \item{Team size: Kenna and Berche~\cite{2012_r.kenna} found that increasing team size beyond a certain point does not improve team quality. Therefore, team sizes should be capped.}
  \item{Worker workload: It is unrealistic for a worker to take on multiple complex tasks simultaneously. Therefore, each worker should be limited to handling only one task at a time.}
\end{itemize}
This study establishes these as the requirements for the team formation problem.
Applying the team formation algorithm that generates a single team for each task sequentially, in turn, may result in teams that do not satisfy the requirements of tasks to be processed later, or in teams that are less compatible with each other.
Therefore, an algorithm capable of concurrently forming teams for multiple tasks is required.

The problem of forming multiple teams for complex tasks concurrently through crowd workers with specialized abilities is referred to as the multi-team formation problem.
The primary aim of this study is to represent this formation problem using mathematical formulas and to analyze the outputs through a straightforward approach.
To solve this problem, we propose a heuristic algorithm that forms teams with optimized compatibility for multiple tasks concurrently while satisfying all requirements.

The properties of the outputs of the proposed algorithm were analyzed through simulation experiments, which revealed how the team's compatibility was influenced by eight simulation parameters and hyperparameters.
Simulation experiments were performed using a synthetic dataset.
Some information, such as the weights of social network edges, worker skill levels, hiring costs, and task skill requirements, was generated from a Poisson distribution, with its parameters included in the experimental parameters.
In the experiments, the objective function, which outputs the team's compatibility for an input vector comprising each experimental parameter, was approximated through Gaussian process regression (GPR) with the observed simulation outcomes.
In addition, Bayesian optimization was employed to determine the experimental conditions necessary for regression.

The main contributions of this study are summarized as follows:
\begin{itemize}
  \item{We formulate a crowdsourcing team formation problem that simultaneously considers skill requirements, budget, team size, and worker compatibility.}
  \item{To solve the problem, we propose a heuristic algorithm that combines a SAT solver to generate feasible initial solutions with simulated annealing for optimization.}
  \item{Through extensive simulations, we analyze the characteristics of the proposed model with respect to various parameters, and demonstrate that our algorithm consistently outperforms hill climbing as a baseline.}
\end{itemize}

%% file: tex/section2.tex
\section{Related work}
There have been two types of applied research on team formation: focusing on applications within organizations and in crowdsourcing.
In team formation within general organizations, all workers are typically required to be assigned to one of the teams, whereas in crowdsourcing team formation, a proportion of workers are selected from a larger worker pool.
In addition, many studies on crowdsourcing team formation have considered factors such as hiring cost and skill levels of workers.

Crowdsourcing often involves tasks where individual workers operate independently, and many applications leverage the collective power of the crowd.
Recent studies have addressed the team formation problem in social networks, which is similar to our mathematical problem.
Furthermore, some studies have applied the team formation problem to crowdsourcing.

\subsection{Crowdsourcing}
Crowdsourcing offers cost-effective, high-quality results by securing labor from an unspecified number of people~\cite{2016_h.kashima}.
Numerous services and systems have emerged or evolved thanks to the development of the Internet.
For example, reCHAPCHA is a crowdsourcing system for transcribing scanned book images to verify whether a user is a human being in specific situations, such as authentication.
Foldit is a system that predicts protein structures by having people play an online puzzle game~\cite{2011_f.khatib}.
Amazon Mechanical Turk and Upwork are marketplaces that connect task outsourcers with willing workers.

\subsection{Team formation problem on social networks}
Many studies have tackled the team-formation problem by evaluating the compatibility of a team with a social network.
Lappas et al. were the first to address the team-formation problem in social networks~\cite{2009_t.lappas}.
They modeled real-life relationships, such as those between bosses and co-workers, in social networks with edge weights as communication costs.
In their study, the communication distance between workers was defined as the sum of the edge weights along the shortest paths in the social network.
They also defined two objective functions for the team formation problem: (1) the diameter of the subgraph representing the largest distance between any two workers in the team; (2) the sum of the edge weights of the minimum Steiner tree formed on the subgraph by the team's nodes.
Subsequent studies~\cite{2010_c.t.li,2012_a.gajewar} proposed a new measure called {\it density}.
This measure is adopted in our study and is explained in the next section.
Awal and Bharadwaj introduced a new measure of team compatibility, incorporating individual knowledge levels and the ability to work together as a group~\cite{2014_g.k.awal}.
Juang et al. proposed a measure of team compatibility for the total distance from the project leader in a social network~\cite{2013_m.c.juang}.
Selvarajah et al. introduced an index integrating explicit ratings among workers, worker profile proximity, and psychological compatibility~\cite{2021_k.selvarajah}.

Lappas et al.~\cite{2009_t.lappas} included 
a requirement for team members to have certain skills in their problem formulation.
Some studies used the number of workers with certain skills as a task requirement~\cite{2010_c.t.li,2015_c.t.li}.
Other studies assigned skill levels to workers and imposed requirements on the total skill level within the team~\cite{2010_c.dorn,2011_f.farhadi,2011_c.dorn}.
In addition, some studies considered worker remuneration for a project and incorporated a budget requirement into the team formation problem~\cite{2012_m.kargar,2013_m.kargar}.
Anagnostopoulos et al. proposed an algorithm that the number of tasks performed by an individual worker, with team compatibility as a constraint~\cite{2012_a.anagnostopoulos}.
Conversely, Majumder et al. proposed an algorithm that optimizes team coordination while limiting the number of tasks a worker can undertake~\cite{2012_a.majumder}.
Rangapuram et al. proposed a team formation algorithm that limits team size to prevent team performance from degrading as the number of team members increases beyond a certain number~\cite{2013_s.rangapuram}.

Some studies have addressed the problem of limiting task duration~\cite{2013_y.yang} or considering geographic proximity~\cite{2020_l.chen}.
In addition, some studies have proposed worker clustering algorithms: a study \cite{2018_k.selvarajah} reduces the computational complexity, while other studies \cite{2014_g.k.awal,2019_k.selvarajah} focus on evolutionary approaches.

\input{table/related_work.tex}

\subsection{Team formation on collaborative crowdsourcing}
Table \ref{tbl:related_work} presents a summary of previous studies on team formation problems in collaborative crowdsourcing.
These studies have addressed the problem of forming teams of crowd workers within social networks.

Many studies on the team formation problem in crowdsourcing incorporate budget constraints due to the nature of crowdsourcing, yet few consider team compatibility.
Rahman et al. addressed the problem of minimizing team coordination metrics, such as maximum distance and total distance within the team, assuming that the distances between all workers are quantified~\cite{2015_h.rahman}.
They partitioned teams to ensure that the number of members per team did not exceed a predefined limit.
Sun et al. proposed an algorithm that first determines a centrality leader and then forms teams while limiting the total distance to that leader~\cite{2016_y.sun}.
Yamamoto and Okamoto introduced a team formation system designed for projects with multiple tasks~\cite{2023_r.yamamoto}.
This system forms teams with optimized compatibility for each task while adhering to the same budget.

Yadav et al. proposed a method for simultaneously applying a team formation algorithm to multiple tasks~\cite{2017_a.yadav}.
In their approach, the objective function minimizes the total hiring costs of workers.
Therefore, their problem formulation differs from our study, which focuses on optimizing team compatibility.

Some studies have investigated the following possibilities: crowd workers may incorrectly report information such as their skill levels and hiring costs~\cite{2015_q.liu,2017_w.wang}; workers in a formed team may not always take on tasks~\cite{2017_w.wang}.
Barnabo's method ensures fairness and diversity concerning various individual attributes, such as gender~\cite{2019_g.barnabo}.

\subsection{Quantifying compatibility}
To quantify team compatibility, we must first quantify compatibility between individual workers.
Rahman et al. quantified the compatibility of all worker combinations based on sociodemographic and psychological characteristics~\cite{2015_h.rahman}.
Several previous studies have represented the history of collaboration as a social network and have used a subgraph with team member nodes to quantify team compatibility.
This study also defined team coordination in a similar manner.

Lappas et al. quantified the edge weights in a social network as the communication cost, which is calculated based on the percentage of collaborative projects between workers~\cite{2009_t.lappas}.
The higher the compatibility, the lower the communication cost.
Validation using IMDb reveals a negative correlation between communication cost and team performance~\cite{2018_w.chen}.
Moreover, in crowdsourcing systems, workers are matched arbitrarily, and the percentage of collaborations does not necessarily guarantee compatibility.
Therefore, in this study, worker compatibility should be calculated based on explicit evaluations of one another.

%% file: table/related_work.tex
\begin{table*}[t]
  \centering
  \caption{Studies of team formation problem on collaborative crowdsourcing}
  \label{tbl:related_work}
  \begin{tabular}{lccccc}
    \hline
    & Skill level & Budget & Team size & Multi-task & Team compatibility \\
    \hline
    Rahman et al.~\cite{2015_h.rahman} & $\checkmark$ & $\checkmark$ & $\checkmark$ & & $\checkmark$ \\
    Sun et al.~\cite{2016_y.sun} & & $\checkmark$ & & & $\checkmark$ \\
    Yamamoto \& Okamoto~\cite{2023_r.yamamoto} & $\checkmark$ & $\checkmark$ & & $\checkmark$ & $\checkmark$ \\
    Yadav et al.~\cite{2017_a.yadav} & $\checkmark$ & $\checkmark$ & & $\checkmark$ & \\
    Liu et al.~\cite{2015_q.liu} & & $\checkmark$ & & & \\
    Wang et al.~\cite{2017_w.wang} & & $\checkmark$ & & & \\
    Barnabo et al.~\cite{2019_g.barnabo} & & $\checkmark$ & & & \\
    \hline
  \end{tabular}
\end{table*}

%% file: tex/section3.tex
\section{Proposed model}
This study mathematically defines the multi-team formation problem and proposes a model to solve this problem.

\subsection{Preliminaries}
\subsubsection{Data model}
We assume a set of $n$ workers $U = \{ u_{1}, u_{2}, \cdots, u_{n} \}$ and a set of $m$ tasks $P = \{ p_{1}, p_{2}, \cdots, p_{m} \}$.
Each worker $u \in U$ has skill levels $[s_{1}, s_{2}, \cdots, s_{l}]$ and a hiring cost $c$.
Similarly, each task $p \in P$ requires skill levels $[t_{1}, t_{2}, \cdots, t_{l}]$ and a budget $b$.
All of these values are nonnegative integers.
The $l$ skills encompass all the required skills, and if a worker $u$ lacks the $k$-th skill, then $s_{k}$ for the worker is 0.
For all tasks, $m$ teams $\{ G_{1}, G_{2}, \cdots, G_{m} \} \subset 2^{U}$ are generated by the proposed multi-team formation algorithm.

This study defines a social network representing the relationships among $n$ workers.
The vertex set of the social network is $U$, and the set of edges is $E$.
In crowdsourcing systems, an edge is added between workers as they collaborate, with the edge weight (a non-negative integer), determined by their mutual evaluations. 
Higher edge weights indicate greater compatibility.
The weight of an edge is denoted by the function $w$.
For instance, $w(u, v)$ represents the weight of the edge connecting workers $u$ and $v$ if edge $(u, v) \in E$; otherwise, if edge $(u, v) \notin E$, $w(u, v) = 0$.

\subsubsection{Objective function}
In previous studies~\cite{2009_t.lappas,2011_m.kargar,2012_m.kargar,2013_m.kargar}, the following four communication cost functions defined for team $G$ have proposed as objective functions for the team formation problem:
\begin{itemize}
  \item{$\text{CC-Diameter}(G)$: The communication cost of the team is defined as the diameter of its subgraph,
    \begin{equation*}
      \displaystyle \text{CC-Diameter}(G) = \max_{\forall u, v \in G} d(u, v).
    \end{equation*}
  }
  \item{$\text{CC-Steiner}(G)$: The communication cost of the team is defined as the total edge weight of the minimum Steiner tree on its subgraph.}
  \item{$\text{CC-SD}(G)$: The communication cost of the team is defined as the total distance between every pair of workers in the team,
    \begin{equation*}
      \displaystyle \text{CC-SD}(G) = \sum_{\forall u, v \in G} d(u, v).
    \end{equation*}
  }
  \item{$\text{CC-LD}(G)$: The communication cost of the team is defined the sum of the distances between from each team member to the leader $u_{L}$,
    \begin{equation*}
      \displaystyle \text{CC-LD}(G) = \sum_{\forall u_{i} \in G, u_{i} \neq u_{L}} d(u_{i}, u_{L}).
    \end{equation*}
  }
\end{itemize}
The distance function $d(u, v)$ for workers $u$ and $v$ in the social network is defined as the sum of the edge weights on the shortest paths between them.

The communication cost functions using the distance on the social network, such as $\text{CC-Diameter}$, $\text{CC-Steiner}$, and $\text{CC-LD}$, assume that shorter distances indicate higher compatibility among workers.
This implies that even if workers have never collaborated, those who are close in the social network are assumed to be more compatible.
However, we found no studies that support correlation between worker compatibility and social network distance in crowdsourcing contexts.
In addition, these functions assume that all workers in the team are connected; this assumption may not hold in sparse graphs.
Therefore, we adopt the density function $f(G)$, which can be computed directly from available edge weights and remains applicable even when the social network is sparse or partially disconnected.

We define a function $f$ for any team $G$ as
\begin{equation*}
  \displaystyle f(G) = \frac{\displaystyle \sum_{\forall u, v \in G} w(u, v)}{\| G \|},
\end{equation*}
which serves as a measure of team compatibility and is referred to as $density$.
In this study, we employ this measure as the objective function of the multi-team formation problem.
This objective function can be calculated only from the weights of the existing edges, even if the subgraph induced by of the team's nodes is disconnected.

\subsection{Problem definition}
This study addresses the problem of concurrently forming teams with optimized compatibility for multiple tasks.
The measure of compatibility within a team is given by the $density$.
Since multiple teams are formed, the objective function is defined as the sum of the $densities$ of all teams.
For each task, the total skill level of the team must exceed the required skill level, and the total employment cost must be within the budget.
Each team must have no more than $K$ members.
Additionally, no worker may be assigned to more than one team.
The mathematical formulation is given as follows:

\begin{align*}
  \text{maximize} \quad & \sum_{j = 1}^{m} \frac{\displaystyle \sum_{\forall i_{1}, i_{2} \in [1, n]} w(u_{i_{1}}, u_{i_{2}}) x_{i_{1j}} x_{i_{2j}}}{\displaystyle \sum_{i = 1}^{n} x_{ij}} \\
  \text{subject}~\text{to} \quad &\sum_{i = 1}^{n} s_{ik} x_{ij} \ge t_{jk}, \quad \forall k \in [1, l],~\forall j \in [1, m], \\
  &\sum_{i = 1}^{n} c_{i} x_{ij} \le b_j, \quad \forall j\in[1,m], \\
  &\sum_{i = 1}^{n} x_{ij} \le K, \quad \forall j\in[1, m], \\
  &\sum_{j = 1}^{m} x_{ij} \le 1, \quad \forall i\in[1, n], \\
  &x_{ij} \in \{0, 1\}.
\end{align*}
Note that $x_{ij}$ denotes whether worker $u_{i}$ is assigned to team $G_{j}$.

\subsection{Proposed multi-team formation algorithm}
The problem of finding a maximal $density$ subgraph whose vertex set size is below a constant $K$ is known to be NP-hard, and its instance set is a subset of the instance set of our defined problem.
Therefore, the problem described above is also NP-hard.

In this study, we propose a heuristic algorithm for the multiteam formation problem.
The basic strategy of the proposed algorithm is {\it Simulated Annealing} (SA).
Due to the nature of SA, it is necessary to determine in advance whether a satisfactory solution exists; if it does, one such solution should be derived.
For this purpose, a Boolean SATisfiability problem (SAT) solver was employed.
Since SAT is NP-complete, the feasibility check has worst-case exponential complexity in the problem size.
In practice, larger $n$, $m$, higher skill requirements $t_{jk}$, and smaller budget $b_{j}$ or team size limit $K$ tend to make the search more difficult.
As noted in Section 4.2, we consider two failure cases as ``no solution'': (i) timeout beyond the cutoff, and (ii) encoding failure due to memory or file-size limits.

\subsubsection{Simulated annealing}
In the proposed algorithm with SA, the probability of transitioning to a neighboring solution with a lower evaluation value decreases over time.
In addition, if the current solution is $G$ and a neighboring solution $G^{\prime}$ (with $f(G^{\prime}) < f(G)$) is selected during the neighborhood search, the probability of transition to $G^{\prime}$ is given by
\begin{equation*}
  \exp \left( \frac{f(G^{\prime}) - f(G)}{T} \right).
\end{equation*}
The temperature $T$ is updated over time according to the equation
\begin{equation}
  T^{\prime} = \alpha T.
\end{equation}
It is noteworthy that $\alpha$ is a hyperparameter of the proposed algorithm.
In our implementation, the initial temperature was set to 10.
The stopping criterion was defined by the overall time limit $L$, where annealing procedure terminates once the allocated time per run is exeeded (details are provided in Section 3.3.3).

In the proposed algorithm, the neighborhood of a team $G$ is defined as the set of candidate solutions that satisfy the constraints on skill levels, budget, and team size.
These neighborhoods are determined from the following sets of teams:
\begin{itemize}
  \item{$N_{1}$: The set of teams formed by removing one worker from team $G$ and adding one worker who does not belong to any team.}  
  \item{$N_{2}$: The set of teams formed by removing two workers from team $G$ and adding one worker who does not belong to any team.}
  \item{$N_{3}$: The set of teams formed by removing one worker from team $G$ and adding two workers who do not belong to any team.}
\end{itemize}
To select a new solution, the proposed algorithm first randomly selects one candidate from each of the three sets, and then randomly chooses one among these candidates.
When the size of $G$ is equal to $K$, only candidates from $N_{1}$ and $N_{2}$ are considered.

To obtain a set of candidate solutions from the neighborhoods, we must verify that every candidate team meets all the requirements.
This verification process has a time complexity of $O(n^{2})$, making it computationally expensive and impractical.
However, since the neighborhood search ultimately requires only one valid solution, the proposed algorithm randomly selects candidate teams until it finds one that satisfies all the requirements.
This procedure is applied to each of $N_{1}$, $N_{2}$, and $N_{3}$, resulting in three types of neighborhood solutions.

\subsubsection{Smoothing for evaluation value}
When evaluating solutions using the {\it density} measure, the evaluation function returns zero for any solution in which no workers in the team are adjacent on the social network.
Consequently, smoothly improving the solution via a nearest neighbor search may be impossible.
Therefore, during the algorithm's execution, a virtual edge is assigned between nonadjacent workers $u$ and $v$ with a weight.
This is defined as
\begin{equation}
  w(u, v) = \left \{
    \begin{array}{ll}
    w(u, v) & (u, v) \in E \\
    \\
    \displaystyle \beta \cdot \frac{1}{p} \cdot \frac{i}{5}, \quad \forall i \in \{0, 1, \cdots, 5\} & (u, v) \notin E,
    \end{array}
    \right.
\end{equation}
where $\beta$ is a hyperparameter, $p$ is the number of edges in the shortest path between $u$ and $v$ in the social network, and $i$ is a nonnegative integer.
The proposed algorithm executes the SA procedure a total of six times, with each run assumed to take the same amount of time.
The counter $i$ is initially set to five and is decremented by one after each iteration of this process.

\subsubsection{Main function}
In addition to the variables defined in this section, the main function of the proposed algorithm accepts as inputs the maximum team size $K$, the hyperparameters in Eq. (1) and (2), the overall time limit for the algorithm $L$, and initial solutions produced by the SAT solver $\{ G_{1}, G_{2}, \cdots, G_{m} \}$.
The pseudocode for the main function is provided in Algorithm \ref{code:main_function} and the explanation is as follows:
\begin{itemize}
  \item{Line 1--3: The initial solutions $G$ obtained from the SAT solver are set as provisional best solutions. $G'$ only stores the best-so-far solutions and does not affect the behavior of the algorithm.}
  \item{Line 4--22: The SA procedure is executed six times. The temperature is reset at the beginning of each run.}
  \item{Line 5: Assign the initial temperature.}
  \item{Line 6--21: For each run, repeat the neighborhood search while lowering the temperature until one-sixth of the total time limit has elapsed.}
  \item{Line 7--19: At each temperature level, repeat the neighborhood search for one six-hundredth of the total time limit.}
  \item{Line 20: After the above time limit has elapsed, lower the temperature.}
  \item{Line 8--13: For each team, conduct the neighborhood search.}
  \item{Line 9: For each team, obtain one neighboring solution. A neighboring solution is a team where one or two workers are replaced, possibly changing the team size.}
  \item{Line 10--12: If a random number drawn from (0, 1) is smaller than the right-hand side of the line 10, the current solution is replaced by the neighboring solution. The higher the temperature, the more likely the algorithm is to accept a worse solution.}
  \item{Line 14--18: Update the best solution. If the sum of $f(G_{j})$ exceeds that of the best-so-far solution, replace it with the current solution.}
\end{itemize}
Optimization is performed for each value of $i$ in Eq. (2), and this process is executed six times in total; therefore, the time limit per run is set to $L / 6$.
The temperature in the SA procedure is updated 100 times per optimization, so time allocated for the neighborhood search at each temperature is set to $L / 600$.
Neighborhood searches are conducted separately for each team, and then each solution is updated.
The probability of updating a solution extracted from the neighborhoods is given by
\begin{equation*}
  \exp \left( \frac{f(G^{\prime \prime}_{j}) - f(G_{j})}{T} \right).
\end{equation*}
A value $rand(0, 1)$ is drawn from a uniform distribution in over the interval $[0, 1]$.
Once all solution updates have been completed, the best solutions are selected and updated.

\input{code/main.tex}

%% file: code/main.tex
\begin{algorithm}[t]
  \caption{Main function}
  \label{code:main_function}
  \begin{algorithmic}[1]
  \Require wokers set $U$, edges set $E$, tasks set $P$, max team size $K$, hyperparameters $\alpha, \beta$, time limit $L$, initial solutions $\{ G_{1}, G_{2}, \cdots, G_{m} \}$
  \Ensure optimized solutions $\{ G^{\prime}_{1}, G^{\prime}_{2}, \cdots, G^{\prime}_{m} \}$
    \ForAll{$j \gets [1, m]$} 
      \State $G^{\prime}_{j} \gets G_{j}$
    \EndFor
    \ForAll{$i \gets [5, 4, \cdots, 0]$} 
      \State $T \gets 10$
      \Repeat
        \Repeat
          \ForAll{$j \gets [1, m]$}
            \State $G^{\prime \prime}_{j} \gets$ a solution extracted from neighborhood
            \If{$rand(0, 1) < \exp((f(G^{\prime \prime}_{j}) - f(G_{j})) / T)$}
              \State $G_{j} \gets G_{j}^{\prime \prime}$
            \EndIf
          \EndFor
          \If{$\sum_{j = 1}^{m} f(G_{j}) > \sum_{j = 1}^{m}  f(G^{\prime}_{j})$}
            \ForAll{$j \gets [1, m]$} 
              \State $G^{\prime}_{j} \gets G_{j}$
            \EndFor
          \EndIf
        \Until{time exceeds $L / 600$} 
        \State $T \gets \alpha \cdot T$
      \Until{time exceeds $L / 6$}
    \EndFor
  \end{algorithmic}
\end{algorithm}

%% file: tex/section4.tex
\input{table/parameter.tex}

\section{Experimental evaluation}
In this study, we conducted a simulation experiment, with three main objectives:
\begin{enumerate}
  \item{To determine the relationship between the experimental parameters and the evaluation values of the proposed algorithm's outputs. In particular, we aim to identify which parameters have a strong influence and how they depend on one another. This information can help establish the system conditions (e.g., the number of registered workers) under which the proposed algorithm is justified.}
  \item{To investigate the relationship between the hyperparameters and the evaluation values in the proposed algorithm. In other words, we explore how the hyperparameters should be adjusted under various experimental conditions.}
  \item{To validate whether the SA procedure and the smoothing technique positively contribute to the optimization performance of the proposed algorithm.}
\end{enumerate}

\subsection{The synthetic dataset}
We could not find real-world datasets on team formation using crowdsourcing.
Therefore, in this study, we employ a stochastically generated synthetic dataset.
Table \ref{tbl:parameter} lists the experimental parameters used to generate the synthetic dataset.
For each combination of experimental parameters, an optimization problem was generated with the following conditions:
\begin{itemize}
  \item{Social network generation: A social network is generated using the LFR benchmark~\cite{2008_a.lancichinetti}, with $n$ representing the number of workers and $k$ denoting the degree mean of each node.}
  \item{Edge weight assignment: Each edge in the social network is assigned an integer weight between 1 and 5, randomly sampled from a Poisson distribution with a mean of 3.}
  \item{Skill types and worker skills: This experiment assumes 20 skill types. The number of skills possessed by each worker is randomly sampled from a Poisson distribution with a mean of 5.}
  \item{Worker skill levels: Each worker's skill level, an integer between 1 and 9, is randomly sampled from a Poisson distribution with a mean of 3.}
  \item{Employment cost: A worker's employment cost is randomly sampled from a Poisson distribution with the mean equal to the sum of the worker's skill levels.}
  \item{Task skill requirements: For each of the $m$ tasks, the number of required skills is randomly sampled from a Poisson distribution with a mean of $m_{SN}$, ensuring that the mean of the required skill counts across all tasks is as close as possible to $m_{SN}$.}
  \item{Required skill levels: The required levels of these skills are randomly sampled from a Poisson distribution with a mean of $m_{SL}$.
  Similarly, the mean of all skill levels is as close as possible to $m_{SL}$.}
  \item{Task budget: The budget for a task is defined as the sum of its required skill levels for each task plus a non-negative integer $b^{\prime}$; that is, the budget for the task $j$ is given by $\displaystyle \sum_{k = 1}^{20} t_{jk} + b^{\prime}$.}
\end{itemize}
In this experiment, an optimization problem is generated from a combination of experimental parameters, and solving that problem yields a single evaluation value.
In this way, a set of experimental parameter-evaluation value pairs can be obtained.

\input{table/gpyopt_settings.tex}

\subsection{Experimental methodology}
This experiment examines the relationship between the experimental parameters listed in Table \ref{tbl:parameter} and the evaluation values of the teams formed using the proposed algorithm.
Since there eight parameters, performing a grid searching would be time-consuming.
Therefore, this study approximates the distribution of experimental parameter pairs and their corresponding evaluation values using GPR.
Bayesian optimization is employed to select the next set of experimental parameters, with the resulting evaluations serving as input for further optimization.
By iteration this process, a set of experimental parameter-evaluation value pairs can be efficiently obtained.
In this experiment, the GPR and Bayesian optimization modules were implemented using {\it Gpy} and {\it GpyOpt}, respectively; the parameters used for {\it GpyOpt} are listed in Table \ref{tbl:setting}.
When the acquisition type is LCB, a larger acquisition weight biases the selection of experimental parameters toward exploration, while a smaller weight favors optimization.
In order to improve the accuracy of the regression, the acquisition weight was set to a relatively high value (the default is two).

In the SAT solver, the constraint satisfaction problem is solved before the generated synthetic data are input into the proposed algorithm.
In this study, we employed {\it sugar} \cite{2008_n.tamura} as a solver that encodes the constraint satisfaction problem to a Boolean CNF formula, which is then solved using the external SAT solver.
For the external solver, {\it minisat}\footnote{Niklas E\'{e}n, Niklas S\"{o}rensson: ``The MiniSat Page,'' \url{http://minisat.se/}, (2025-1-2)} was chosen for this experiment.
The program to run {\it sugar} was implemented using the officially distributed {\it sugar-java-api}.
Solving SAT problems with a large search space but a small solution space is extremely time-consuming.
Therefore, if the computation takes more than one hour, we assume that no solution exists.
In addition, if the variable space or the number of constraints becomes excessively large, these cases may exceed the upper limit of the file size to be encoded, making it impossible to run the solver.
Such experimental parameters are treated as cases with no solution.

After the solver generates an initial solution, the proposed algorithm is executed.
A grid search is conducted over the hyperparameters $\alpha$ and $\beta$; specifically, the algorithm is run for all hyperparameter pairs with $\alpha \in \{ 0.8, 0.85, 0.9, 0.95 \}$ and $\beta \in \{ 0.5, 1, 1.5, 2 \}$.
The best evaluation results from these 16 combinations are then selected for Bayesian optimization.
A grid search over the hyperparameters is performed, yielding feasible solutions from 16 distinct runs for each set of experimental parameters.
By performing regression on each of these outputs and comparing the results, we can determine which experimental parameters are key and how the hyperparameters should be adjusted.
In this comparative experiment, the output results are compared with those obtained from the hill-climbing method to demonstrate the effectiveness of SA.
Furthermore, to confirm the effectiveness of smoothing, the output results for the case $\beta = 0$ are included in the comparison.
The proposed algorithm was implemented in C++, and all algorithms were executed on a computer with dual Intel Xeon Gold 6132 2.60GHz CPU and 192GB RAM.

%% file: table/parameter.tex
\begin{table*}[t]
  \centering
  \caption{Experimental parameters}
  \label{tbl:parameter}
  \begin{tabular}{ccl}
    \hline
    Parameter & Range & \multicolumn{1}{c}{Description} \\
    \hline
    $n$ & $[100, 1000]$ & The number of workers \\
    $m$ & $[1, 10]$ & The number of tasks \\
    $K$ & $[1, 50]$ & The limit of team size \\
    $k$ & $[5, 30]$ & The degree mean \\
    $m_{SN}$ & $[2, 10]$ & The mean of required skill numbers \\
    $m_{SL}$ & $[5, 45]$ & The mean of required skill levels \\
    $b^{\prime}$ & $[0, 500]$ & The additional budget \\
    $L$ & $\{300, 450, 600\}$ & The time limit of the algorithm [sec] \\
    \hline
  \end{tabular}
\end{table*}

%% file: table/gpyopt_settings.tex
\begin{table}[t]
  \centering
  \caption{GpyOpt parameter settings}
  \label{tbl:setting}
  \begin{tabular}{ll}
    \hline
    Parameter name & Value \\
    \hline
    acquisition\_type & LCB \\
    acquisition\_weight & 5 \\
    kernel & Matern52 (ARD=True) \\
    f & None \\
    de\_duplication & True \\
    \hline
  \end{tabular}
\end{table}

%% file: tex/section5.tex
\section{Result and discussion}
Bayesian optimization yielded results for 4,340 experimental conditions -- i.e., unique combinations of experimental parameters -- of which 1,921 produced feasible optimization problems.
Although regression can be performed using both feasible and infeasible solutions, we conducted regression using only feasible solutions.

\subsection{Analysis of experimental parameters}
\subsubsection{Fixed single parameter cases}
We analyze the relationship between each experimental parameter and its corresponding evaluation value.
For the parameter vector $\boldsymbol{p} = (n, m, K, k, m_{SN}, m_{SL}, b^{\prime}, L)$, one variable is held fixed while the remaining variables are randomly set within the ranges defined in Table \ref{tbl:parameter}.
For these randomly set variables, 100,000 samples are generated (with replacement). 
For each combination of a fixed value and a generated sample, we compute $\boldsymbol{\theta} = g(\boldsymbol{p} | P, \boldsymbol{y})$, where $\boldsymbol{\theta} = (\mu, \sigma^{2})$ denotes the mean and variance of the estimated evaluation value, $g$ is a GPR function, $P \in \mathbb{N}^{1921 \times 8}$ is the matrix of previously observed evaluation parameters corresponding to the 1,921 feasible optimization problems, and $\boldsymbol{y} \in \mathbb{R}_{\geq 0}^{1921}$ is the vector of evaluation value associated with $P$.
This procedure is repeated for all 100,000 samples, and by averaging the results, the mean and variance corresponding to a given fixed value are computed.
The fixed variable is then varied one step at a time over the range specified in Table \ref{tbl:parameter}.
Fig. \ref{fig:x} illustrates the resulting mean and variance; the blue line represents the mean evaluation value, while the light-blue shaded area indicates the 95\% confidence interval.

\input{figure/x.tex}

According to Fig. \ref{fig:x}, we found the following insights:
\begin{itemize}
  \item{(a), (d), (g): The graphs for the number of workers $n$, the degree mean $k$, and the additional budget $b^{\prime}$ exhibit a monotonically increasing trend. The confidence intervals widen when $n$ and $b^{\prime}$ are small, possibly due to the scarcity of feasible solutions for these parameters.}
  \item{(c): The graph for the limit of team size $K$ increases monotonically up to 30 but tends to decrease beyond 40, likely because the disadvantage of a larger search space becomes more significant when $K$ is extremely high.}
  \item{(b), (f): The graphs for the number of tasks $m$ and the mean of the required skill levels $m_{SL}$ generally decrease monotonically.}
  \item{(e): The graph for the mean of the required skill numbers $m_{SN}$ does not exhibit a significant change in the evaluation values. In this experiment, 20 skill types were assumed, with workers possessing an average of three skills, suggesting that workers complement each other well in meeting the required skill levels.}
  \item{(h): The graph for the time limit of the algorithm $L$ also shows minor variations, and in most cases, the annealing method converges earlier to a near-optimal solution.}
\end{itemize}

\subsubsection{Fixed two parameters cases}
In the previous section, we analyzed the effect of each experimental parameter on the estimated evaluation value by observing how the values change when one parameter is varied while all other parameters are marginalized.
However, in practice, these parameters are not always independent; the evaluation values may become extremely low depending on the combination of the parameter values.
For example, Fig. \ref{fig:x0_x1} shows that the evaluation values estimated by GPR for parameter $n$ vary with different values of $m$, with particularly significant fluctuations observed when $n$ is small.
Note that, except for $m$ and $n$, all estimated evaluation values in Fig. \ref{fig:x0_x1} were computed using the same method as in Fig. \ref{fig:x}.

\input{figure/x0_x1.tex}

To analyze the independence between any two parameters, $p_{1}$ and $p_{2}$, we introduce the following metrics:
\begin{itemize}
  \item{The {\em variation} due to $p_{1}$ at a fixed $p_{2}$ is defined as the difference between the maximum and minimum estimated evaluation values obtained by varying $p_{1}$. For example, in Fig. \ref{fig:x0_x1}, $p_{1}$ and $p_{2}$ correspond to $m$ and $n$, respectively.}
  \item{The {\em maximum variation} due to $p_{1}$ is defined as the largest variation across all values of $p_{2}$.}
  \item{The {\em average variation} due to $p_{1}$ is defined as the difference between the maximum and minimum mean evaluation values, as indicated by the blue line in Fig. \ref{fig:x}.}
\end{itemize}
In this analysis, except for $p_{1}$ and $p_{2}$, the remaining parameters are marginalized using the same method as in Fig. \ref{fig:x}.
If $p_{1}$ and $p_{2}$ are independent, the variation in $p_{1}$ (measured at any fixed $p_{2}$) remains nearly constant, roughly matching the average variation in $p_{1}$.
Therefore, a significant difference between the maximum variation in $p_{1}$ (as $p_{2}$ varies) and its average variation suggests a potential interaction between $p_{1}$ and $p_{2}$.
Table \ref{tbl:influence} lists, for all pairs of parameters, the differences between the maximum and average variations.
In particular, the entry in the row corresponding to $p_{2}$ and the column corresponding to $p_{1}$ represents the difference between the maximum variation in $p_{1}$ (as $p_{2}$ varies) and the average variations in $p_{1}$.

\input{table/influence.tex}

In Table \ref{tbl:influence}, the parameter pairs with particularly high values are $(K, k)$, $(K, b^{\prime})$, and $(k, b^{\prime})$.
In other words, the parameters $K$, $k$, and $b^{\prime}$ are not independent.
Heatmaps of the estimated evaluation values for these pairs are presented in Fig. \ref{fig:heatmap} (a), (b), and (c).
Fig. \ref{fig:heatmap} shows that when one of the parameters $K$, $k$, or $b^{\prime}$ is small, the evaluation values remain low regardless of the other parameter values.
This occurs because a small $K$ or $b^{\prime}$ limits the number of teams that can be formed, and a small $k$ results in a naturally sparse graph, which in turn leads to low evaluation values.
Therefore, these results suggest that when the team size limit, degree mean, and additional budget are low, the proposed algorithm either fails to form teams that meet the requirements or produces teams with low compatibility.

\input{figure/heatmap.tex}

Table \ref{tbl:influence} indicates that, among the parameter pairs (excluding $K$, $k$, and $b^{\prime}$), the highest values are observed for $(n, m)$ and $(n, m_{SL})$.
The corresponding heatmaps are shown in Fig. \ref{fig:heatmap} (d) and (e).
According to these heatmaps, when $n$ is small and either $m$ or $m_{SL}$ is large, the evaluation value decreases substantially.
This occurs because a large number of tasks or tasks with high skill-level requirements require more workers.

\subsection{Analysis of hyperparameters}
In this experiment, for each of the 1,921 feasible optimization problems, a grid search over the hyperparameters yielded 16 distinct outcomes, each comprising a solution and its corresponding evaluation value.
Subsequently, either $\alpha$ or $\beta$ is held fixed, and the four evaluation values corresponding to the other parameter are averaged.
Finally, as in Fig. \ref{fig:x}, one experimental parameter is varied while the remaining parameters are randomly sampled.

\subsubsection{Fixed $\alpha$ cases}
The results of the $\alpha$ comparison are shown in Fig. \ref{fig:alpha}.
The vertical and horizontal axes represent evaluation values and experimental parameters, respectively.
In addition, the hill-climbing method was applied to every problems that produced a feasible solution, and the estimated evaluation values based on GPR are plotted in Fig. \ref{fig:alpha}.

\input{figure/alpha.tex}

Fig. \ref{fig:alpha} shows that, overall, there is no significant difference in evaluation values among $\alpha = 0.8$, $\alpha = 0.85$, and $\alpha = 0.9$.
The best performance was observed for $\alpha = 0.9$, while the worst was seen at observed for $\alpha = 0.95$.
Alternatively, if the temperature is lowered too slowly or remains high for too long -- resulting in a prolonged high probability of degradation -- a near-optimal solution becomes difficult to achieve.
This discrepancy is especially pronounced when both the number of workers and the number of tasks are large, possibly because a larger solution space requires more time to improve the solution; if the period during which alterations are likely is extended, there will be insufficient time to reach the optimal solution. 
Moreover, the shorter the algorithm's time limit, the greater the discrepancy between the results for $\alpha = 0.95$ and those for the other $\alpha$ values.
Compared to the hill-climbing method, the proposed algorithm generally produced higher evaluation values.
On average, the proposed algorithm achieved evaluation value improvements of at least 16.5\%, 10.2\%, 24.6\%, 9.41\%, 12.2\%, 11.6\%, and 6.84\% for Fig. \ref{fig:alpha} (a)--(g), respectively.

\subsubsection{Fixed $\beta$ cases}
The same analysis was performed for $\beta$, and the results are shown in Fig. \ref{fig:beta}.
For comparison, we computed the evaluation values for all $\alpha$ values with $\beta = 0$.
Note that setting $\beta = 0$ is equivalent to applying no smoothing in the optimization.
Fig. \ref{fig:beta} indicates that the evaluation values are most favorable when $\beta = 0$, and the difference between these values and those obtained for small $\beta$ is trivial.
Therefore, the smoothing process appears to be ineffective.

\input{figure/beta.tex}

%% file: figure/x.tex
\begin{figure*}[t]
  \centering
  \begin{tabular}{cc}
    \includegraphics[width=0.4\hsize]{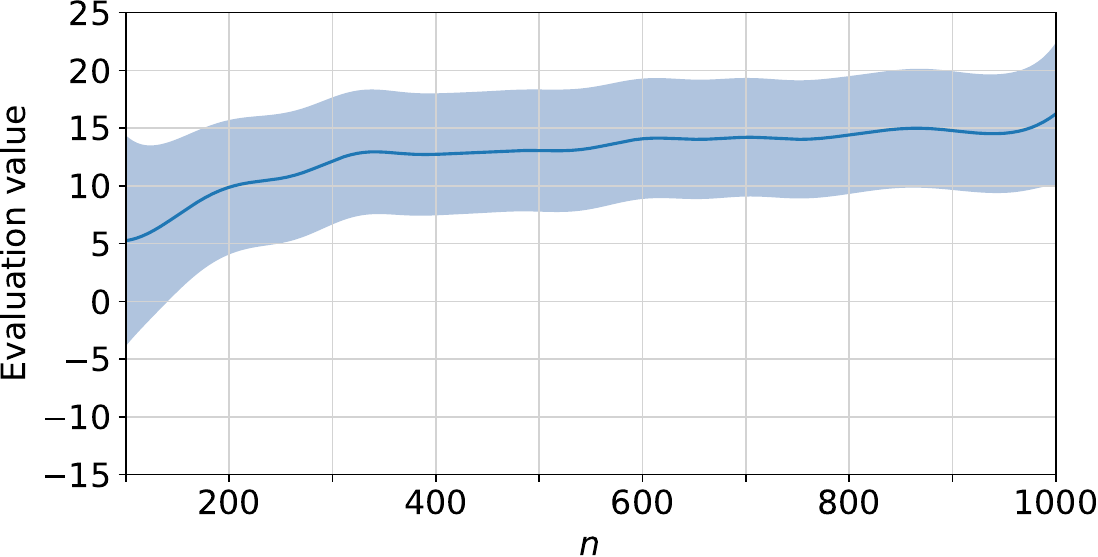} &
    \includegraphics[width=0.4\hsize]{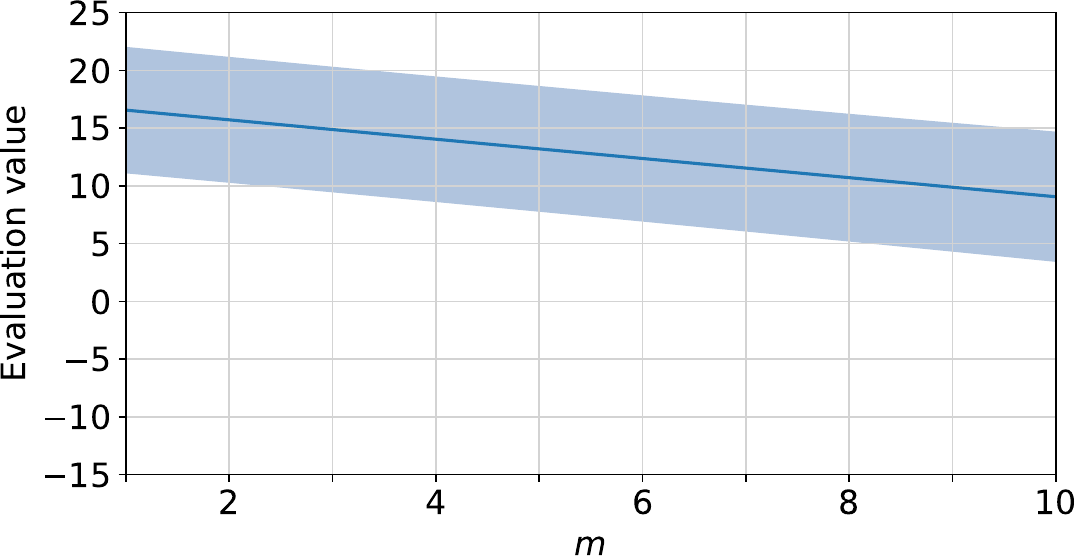} \\
    (a) $n$ & (b) $m$ \\
    \includegraphics[width=0.4\hsize]{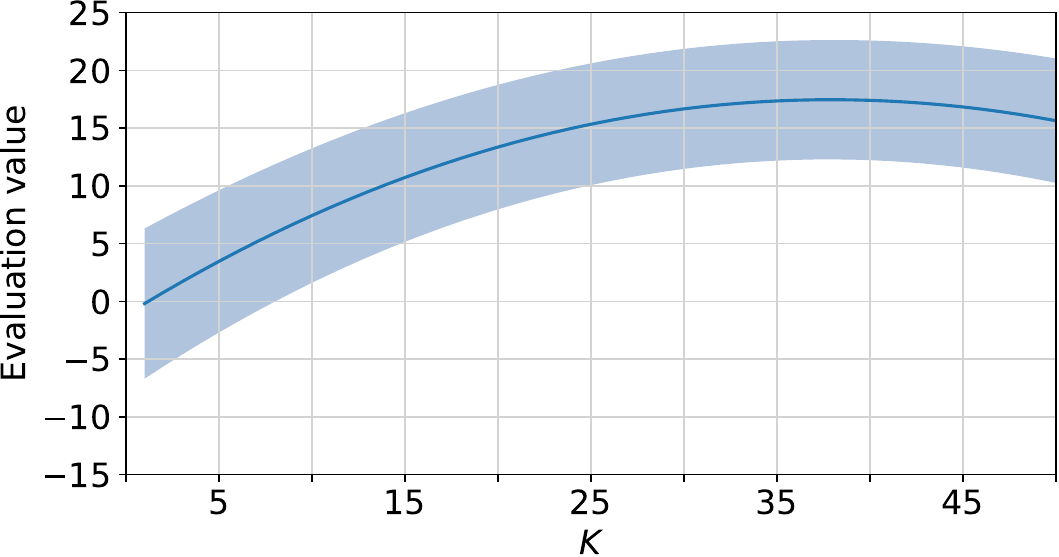} &
    \includegraphics[width=0.4\hsize]{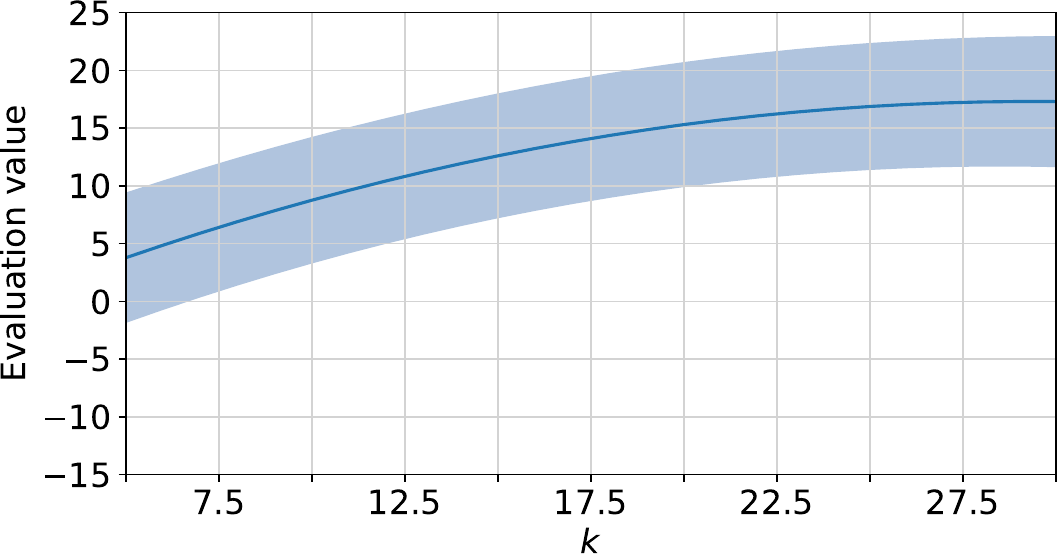} \\
    (c) $K$ & (d) $k$ \\
    \includegraphics[width=0.4\hsize]{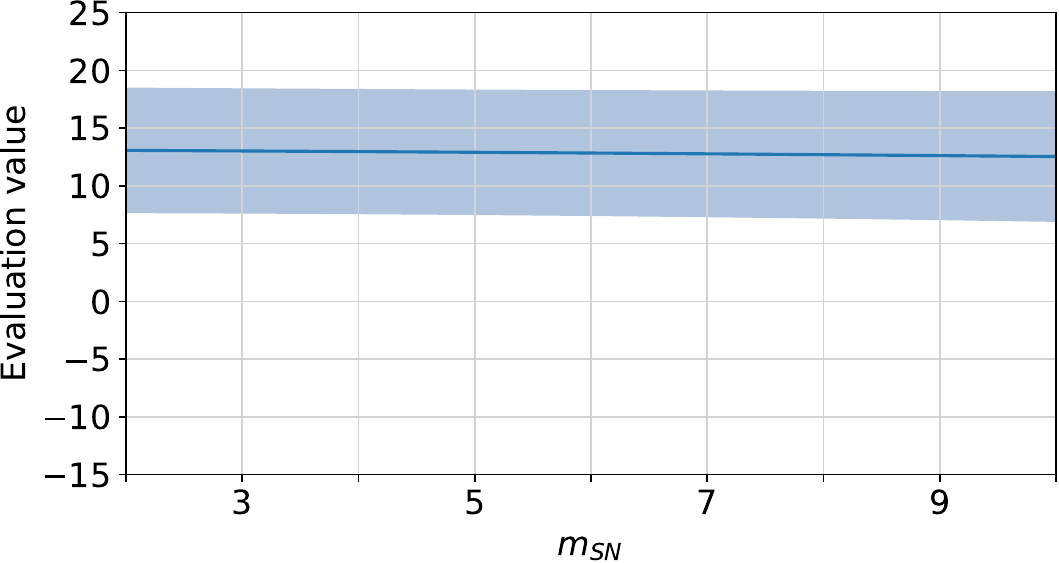} &
    \includegraphics[width=0.4\hsize]{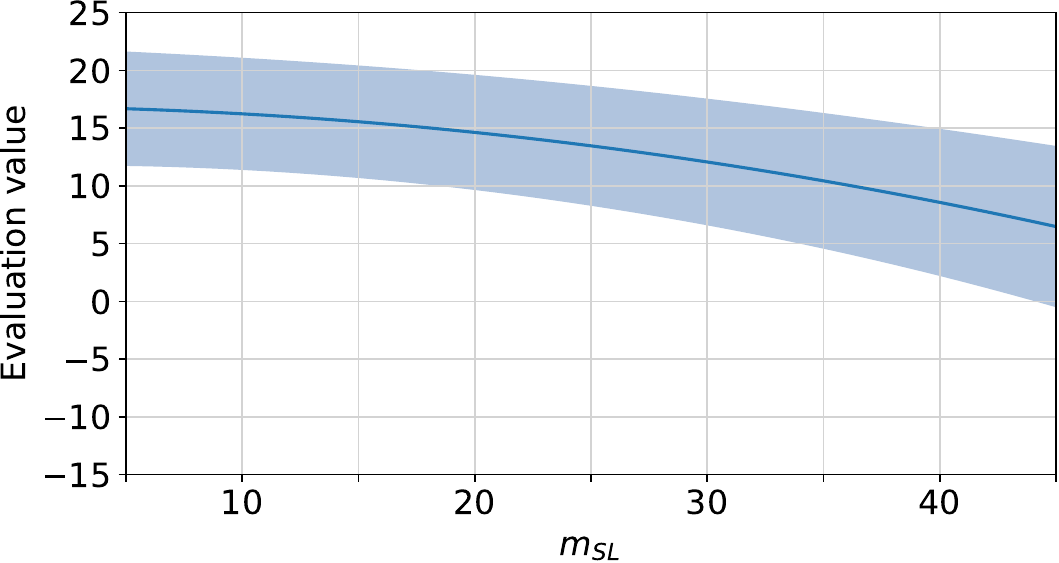} \\
    (e) $m_{SN}$ & (f) $m_{SL}$ \\
    \includegraphics[width=0.4\hsize]{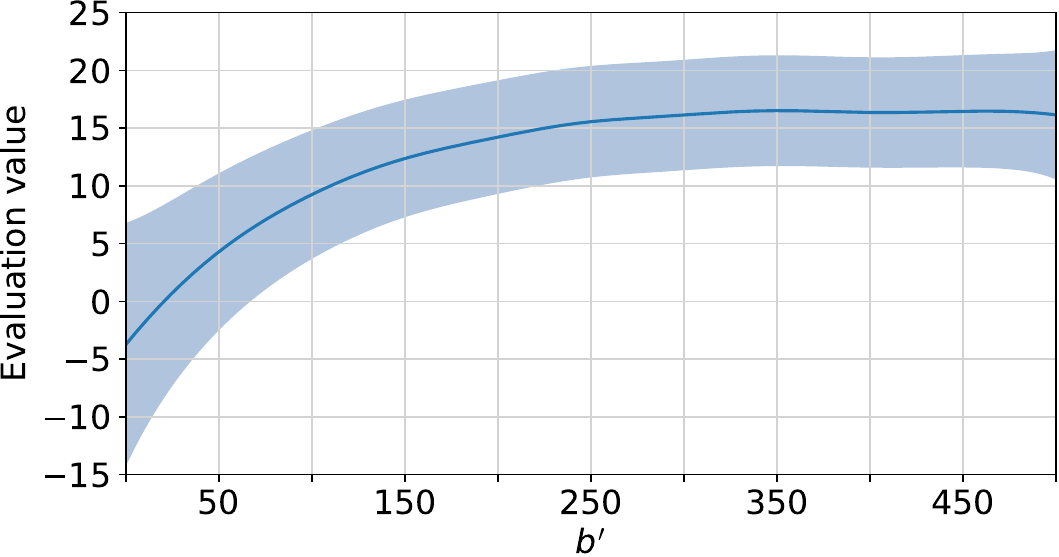} &
    \includegraphics[width=0.4\hsize]{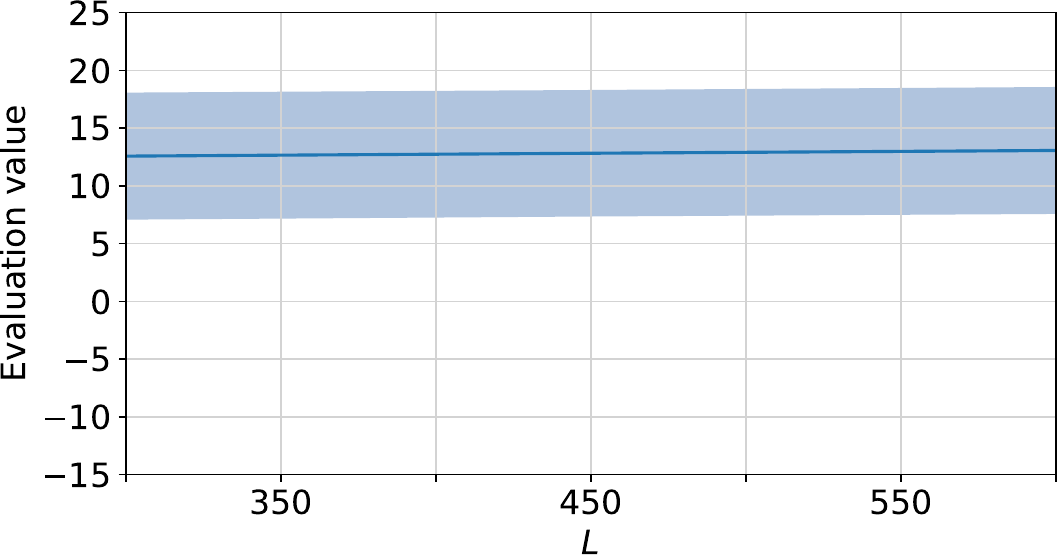} \\
    (g) $b^{\prime}$ & (h) $L$
  \end{tabular}
  \caption{Estimated evaluation values for each experimental parameter}
  \label{fig:x}
\end{figure*}

%% file: figure/x0_x1.tex
\begin{figure}[t]
  \centering
  \includegraphics[width=0.85\hsize]{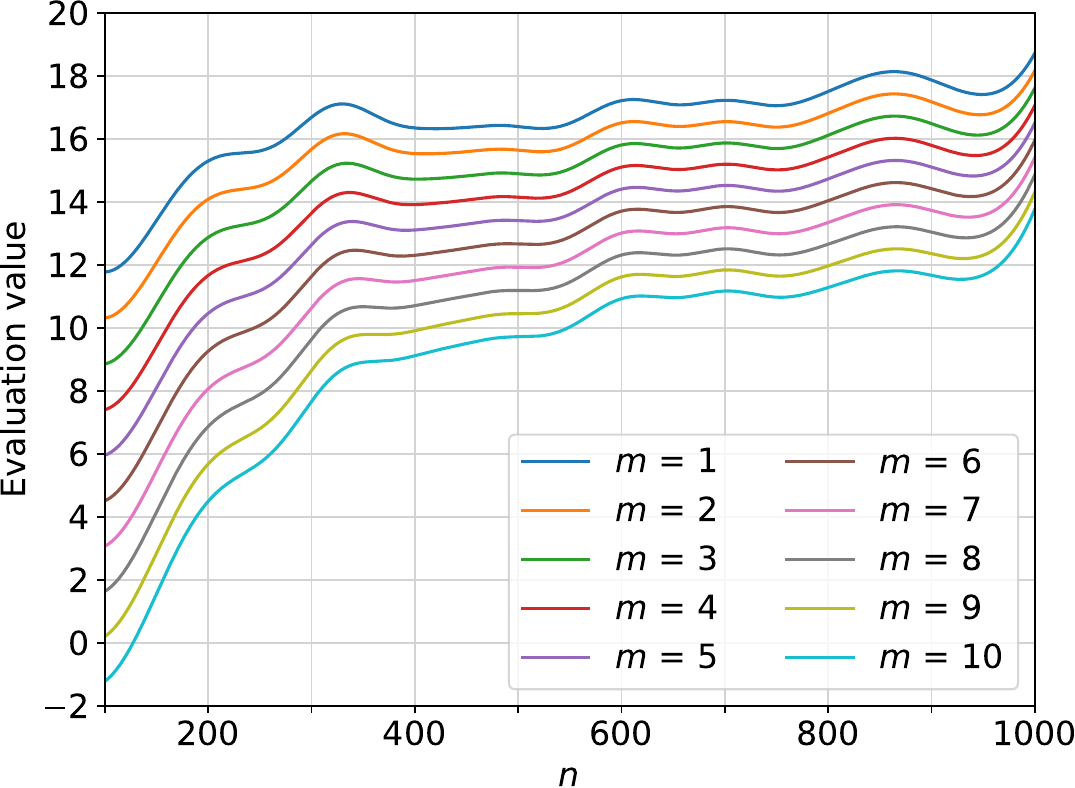}
  \caption{Estimated evaluation values for parameter $n$ vs. $m$}
  \label{fig:x0_x1}
\end{figure}

%% file: table/influence.tex
\begin{table*}[t]
  \caption{Difference between maximum and average variations for each parameter pair (row: $p_{2}$, column: $p_{1}$)}
  \label{tbl:influence}
  \centering
  \begin{tabular}{c|cccccccc} 
    \hline
    & $n$ & $m$ & $K$ & $k$ & $m_{SN}$ & $m_{SL}$ & $b^{\prime}$ & $L$ \\
    \hline
    $n$ & 0.0000 & 5.4968 & 3.9187 & 2.2635 & 1.7379 & 1.4265 & 3.2013 & 0.4146 \\
    $m$ & 4.0000 & 0.0000 & 4.2647 & 3.7021 & 0.3676 & 1.2027 & 3.8679 & 0.0033 \\
    $K$ & 3.9641 & 5.3501 & 0.0000 & 7.7454 & 1.1540 & 2.7514 & 8.6033 & 0.1750 \\
    $k$ & 1.3749 & 3.7234 & 6.4868 & 0.0000 & 1.4495 & 1.0462 & 6.5436 &0.0284 \\
    $m_{SN}$ & 0.8613 & 0.3460 & 0.9020 & 1.4347 & 0.0000 & 0.8693 & 1.2565 & 0.0011 \\
    $m_{SL}$ & 2.2936 & 1.1853 & 2.1980 & 1.0195 & 0.8535 & 0.0000 & 2.2059 & 0.0056 \\
    $b^{\prime}$ & 1.8463 & 3.8112 & 11.789 & 4.9661 & 1.4273 & 3.2042 & 0.0000 & 0.7432 \\
    $L$ & 0.0478 & 0.0024 & 0.1396 & 0.0012 & 0.0117 & -0.0089 & 0.2501 & 0.0000 \\
    \hline
  \end{tabular}
\end{table*}

%% file: figure/heatmap.tex
\begin{figure*}[t]
  \centering
  \begin{tabular}{cc}
    \includegraphics[width=0.35\hsize]{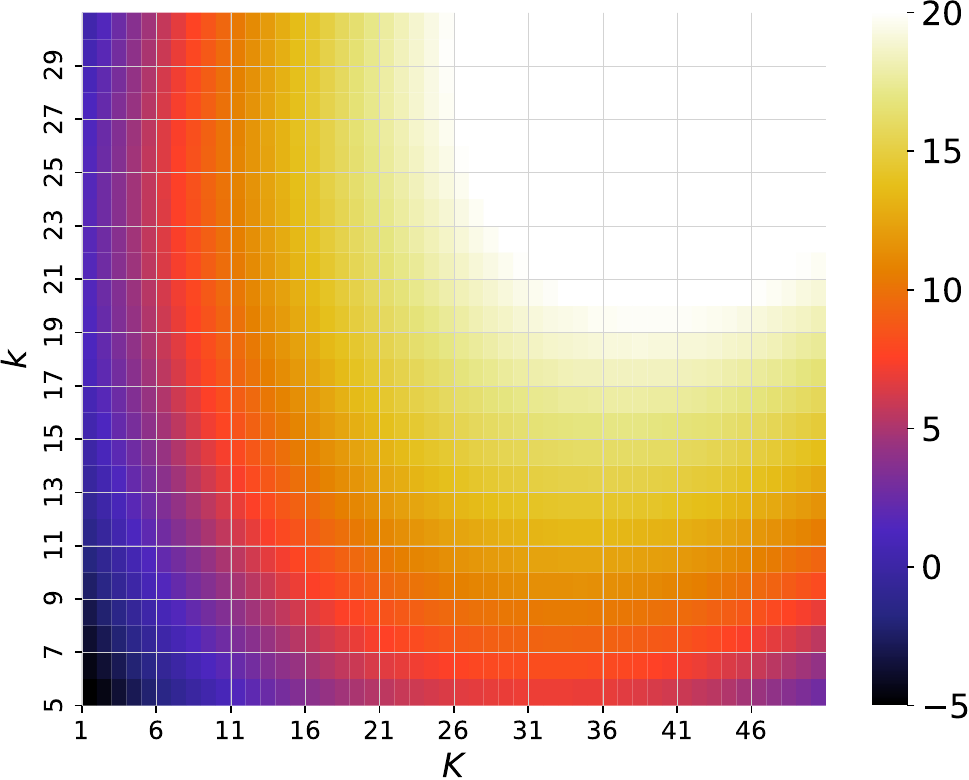} &
    \includegraphics[width=0.35\hsize]{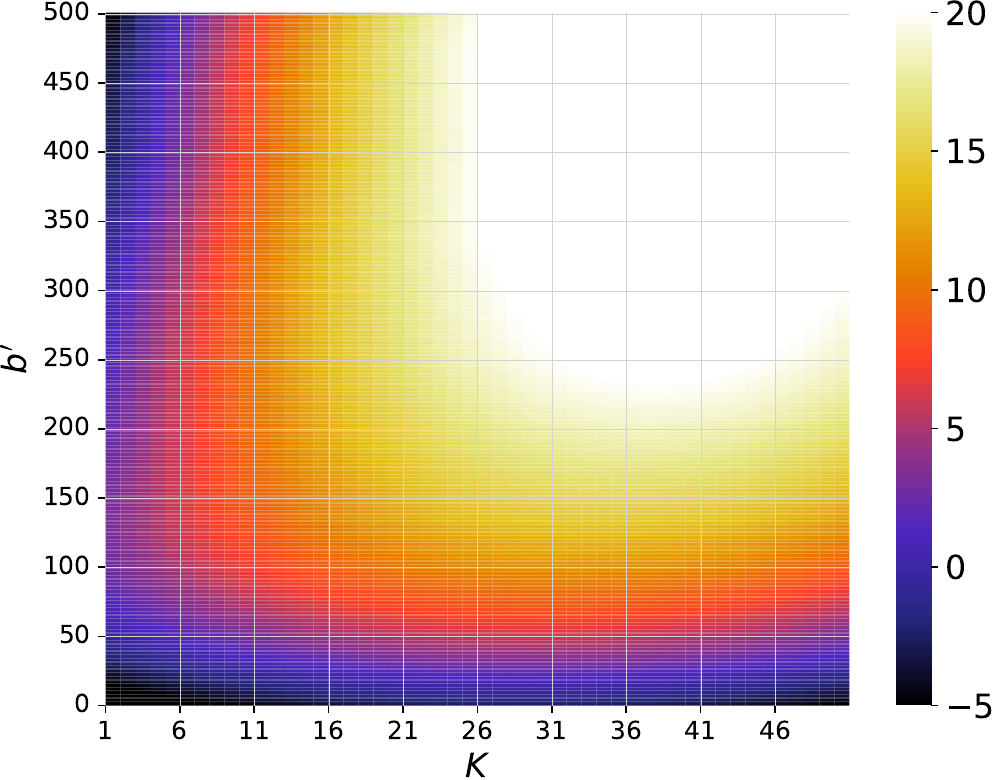} \\
    (a) $K$ and $k$ & (b) $K$ and $b^{\prime}$ \\
    \includegraphics[width=0.35\hsize]{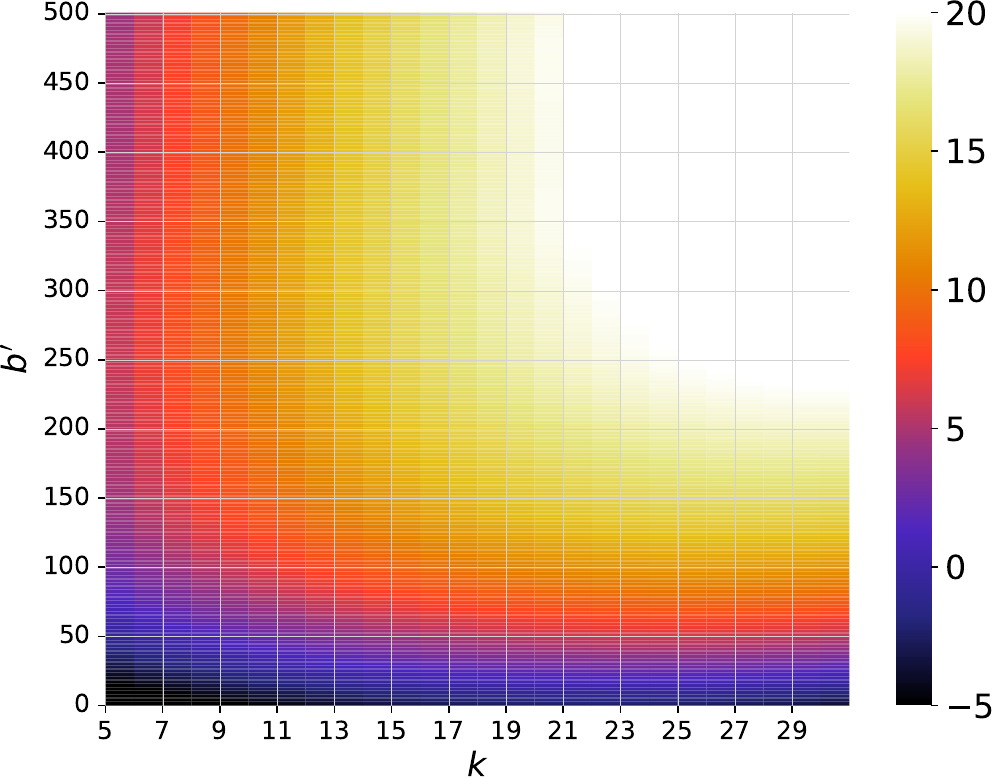} &
    \includegraphics[width=0.35\hsize]{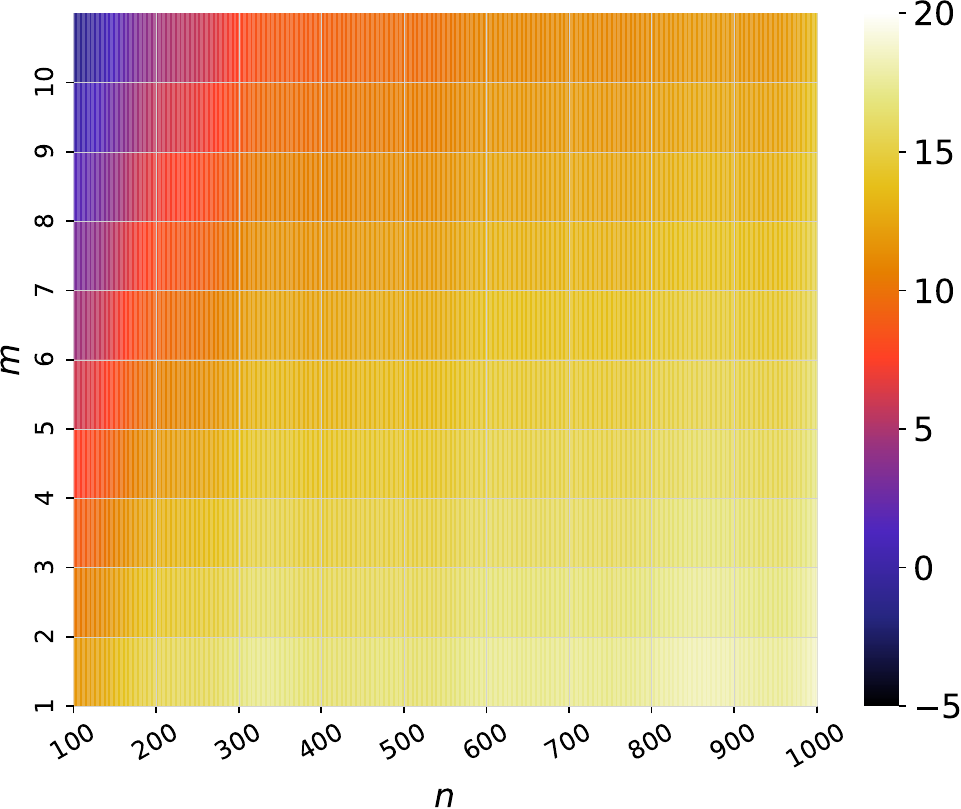} \\
    (c) $k$ and $b^{\prime}$ & (d) $n$ and $m$ \\
    \includegraphics[width=0.35\hsize]{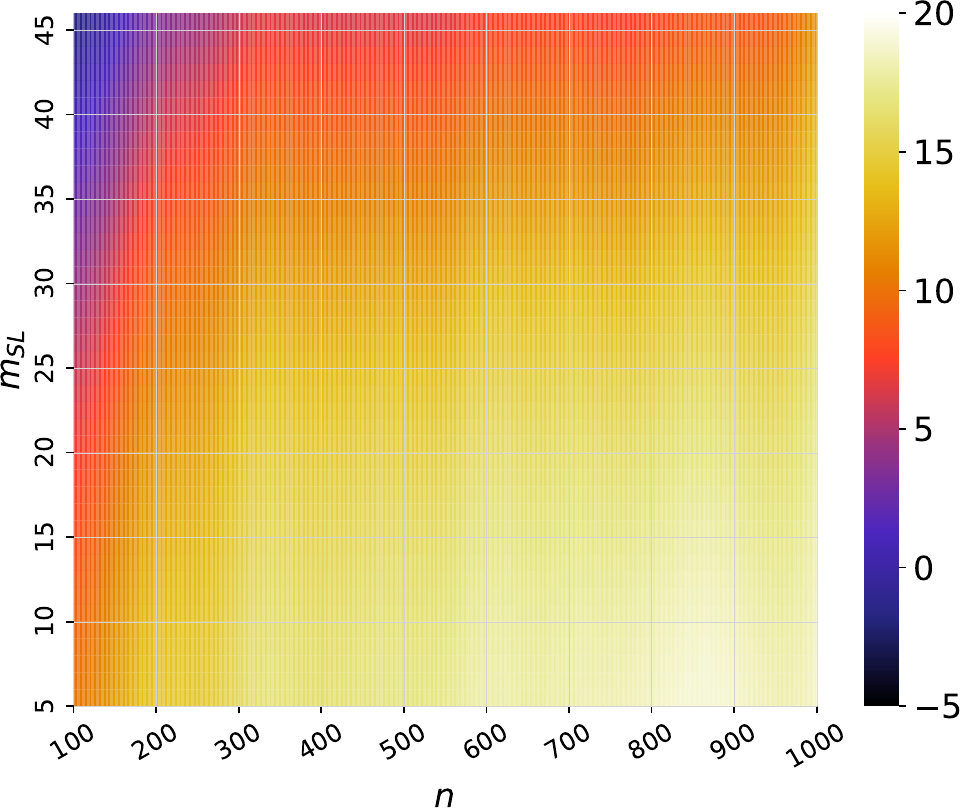} & \\
    (e) $n$ and $m_{SL}$ &
  \end{tabular}
  \caption{Estimated evaluation values for each parameter pairs}
  \label{fig:heatmap}
\end{figure*}

%% file: figure/alpha.tex
\begin{figure*}[t]
  \centering  
  \begin{tabular}{cc}
    \includegraphics[width=0.4\hsize]{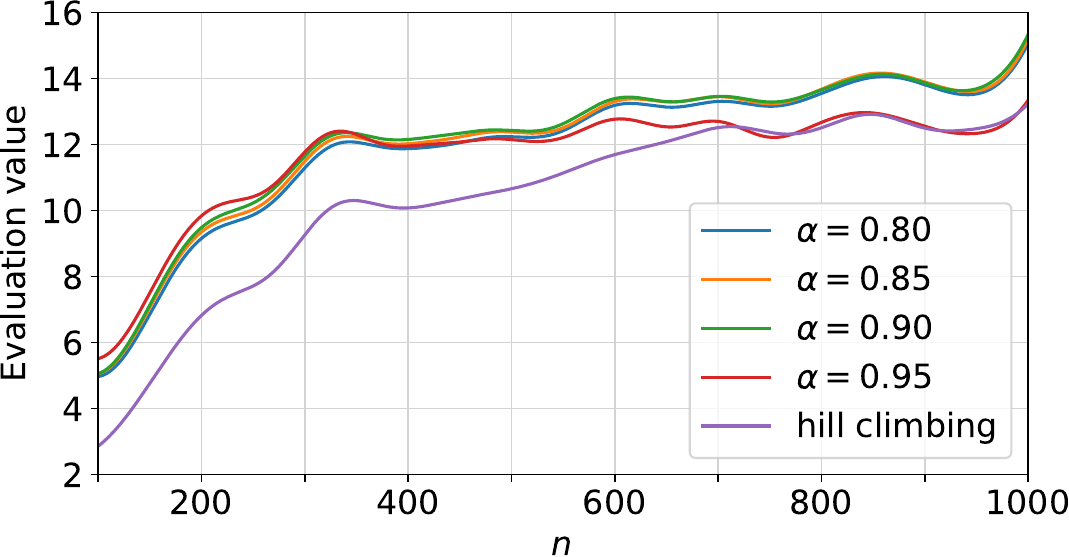} &
    \includegraphics[width=0.4\hsize]{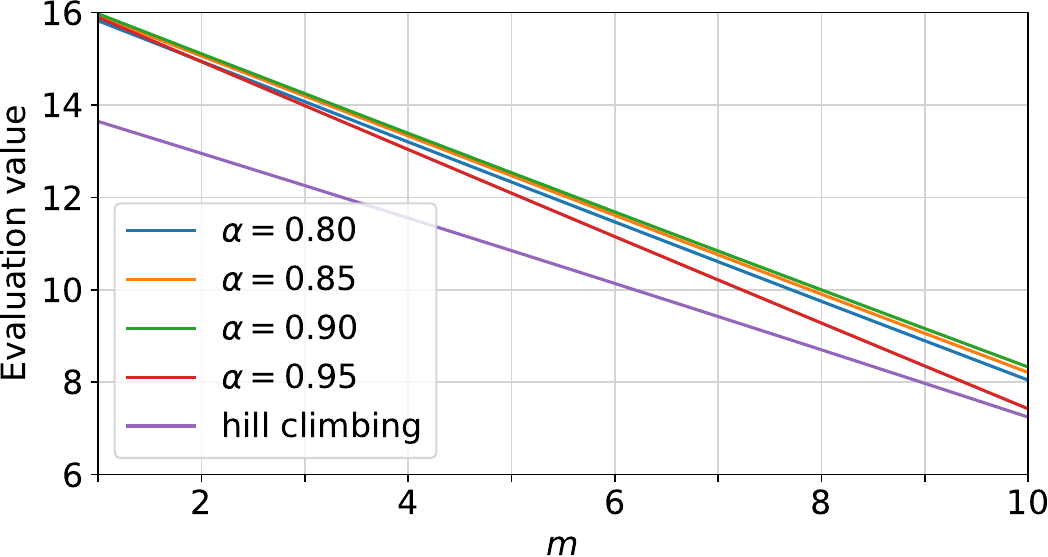} \\
    (a) $n$ & (b) $m$ \\
    \includegraphics[width=0.4\hsize]{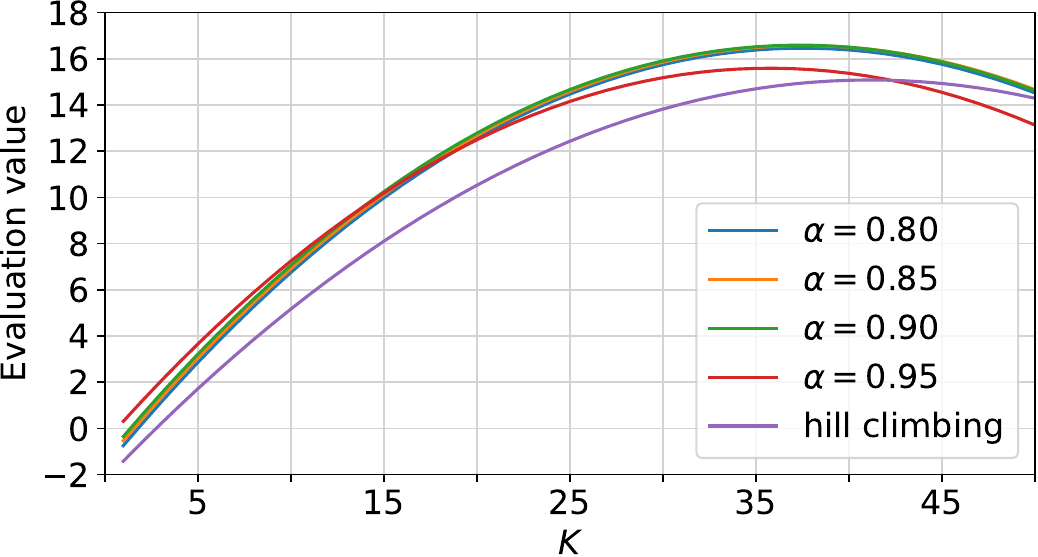} &
    \includegraphics[width=0.4\hsize]{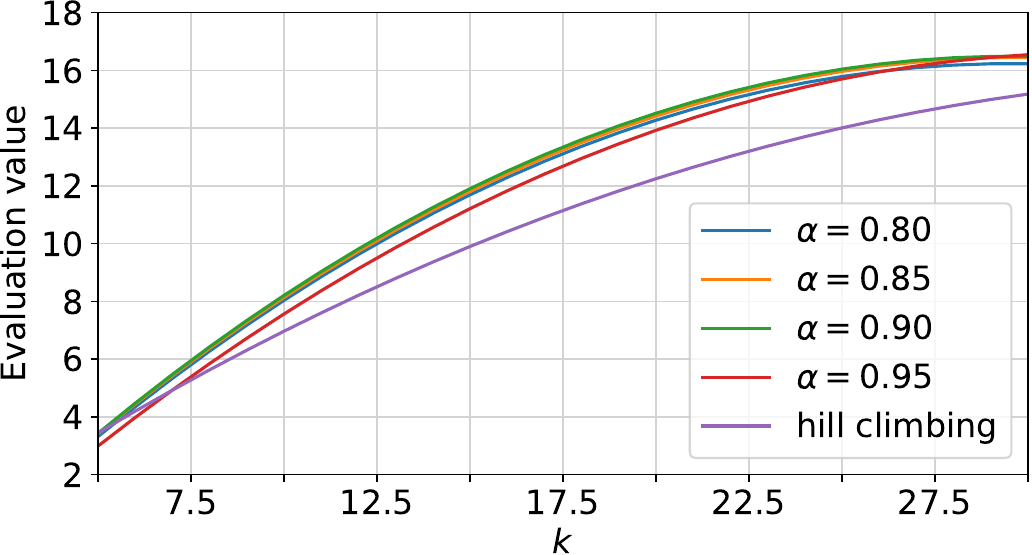} \\
    (c) $K$ & (d) $k$ \\
    \includegraphics[width=0.4\hsize]{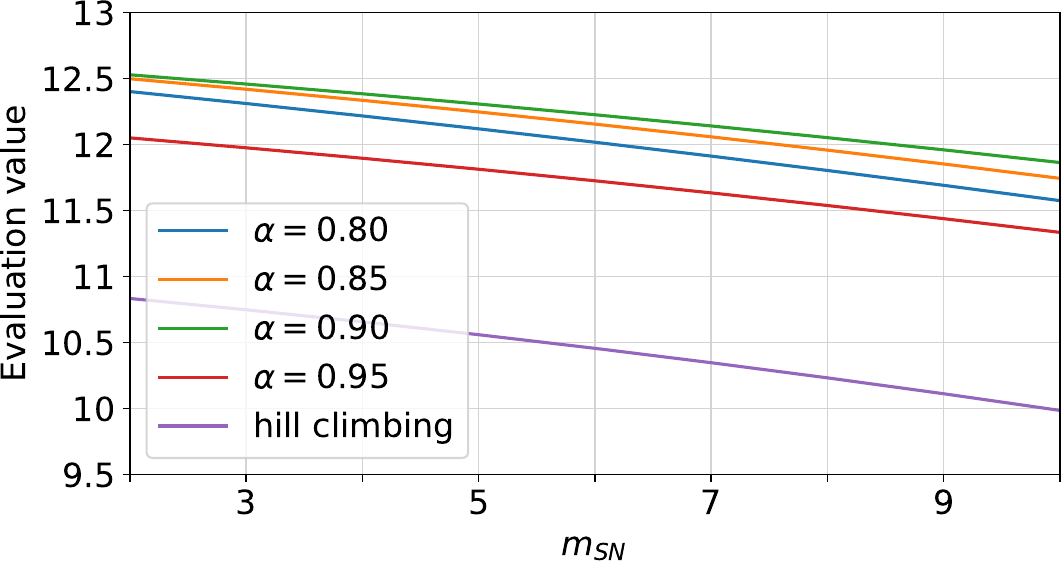} &
    \includegraphics[width=0.4\hsize]{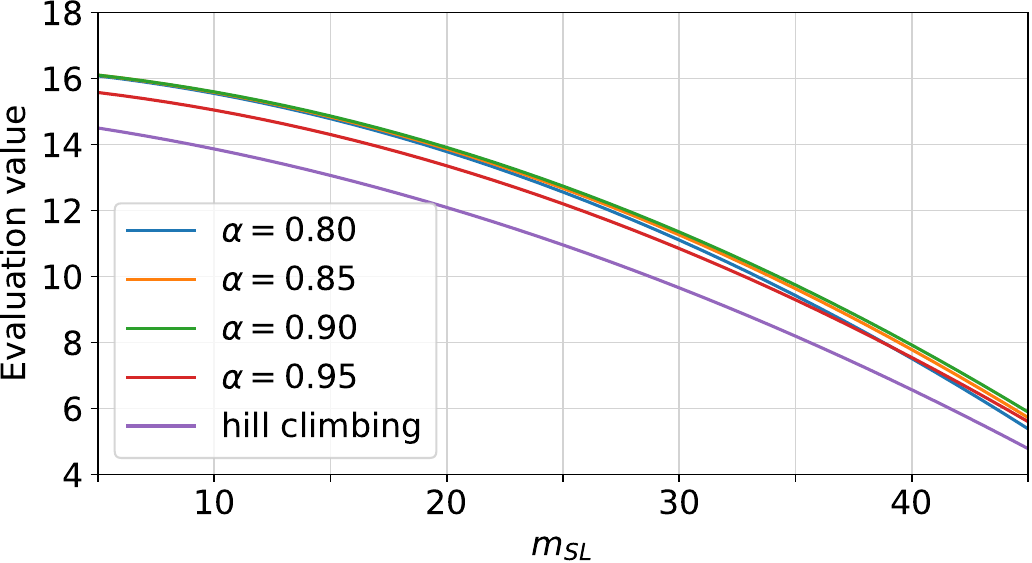} \\
    (e) $m_{SN}$ & (f) $m_{SL}$ \\
    \includegraphics[width=0.4\hsize]{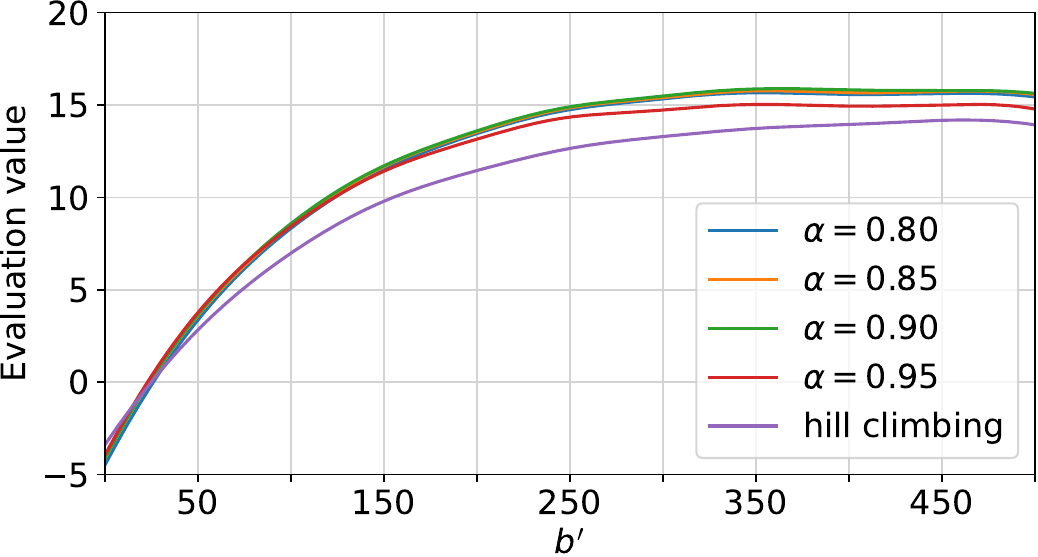} &
    \includegraphics[width=0.4\hsize]{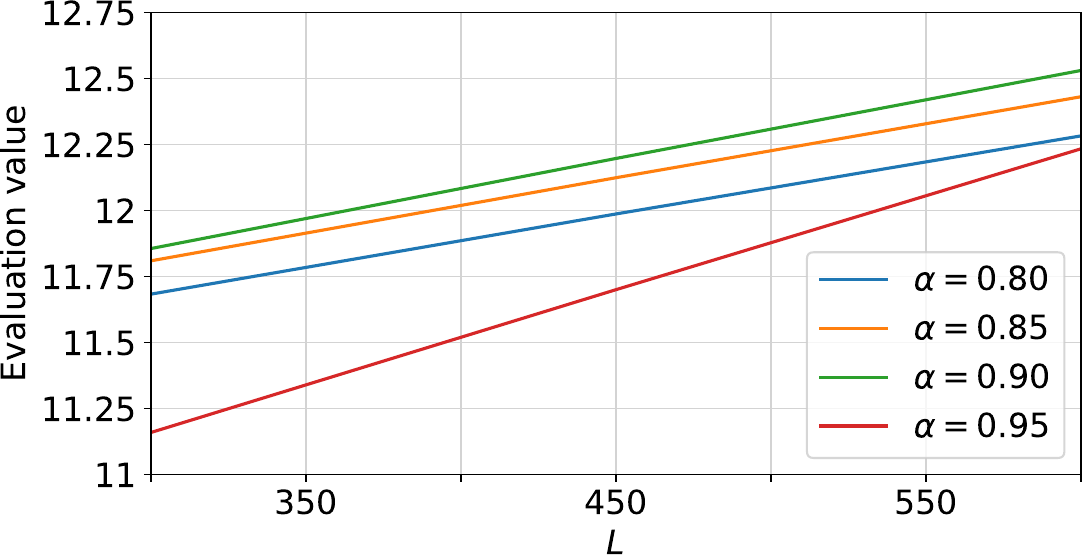} \\
    (g) $b^{\prime}$ & (h) $L$
  \end{tabular}
  \caption{Estimated evaluation values for each experimental parameter at different $\alpha$}
  \label{fig:alpha}
\end{figure*}

%% file: figure/beta.tex
\begin{figure*}[t]
  \centering
  \begin{tabular}{cc}
    \includegraphics[width=0.4\hsize]{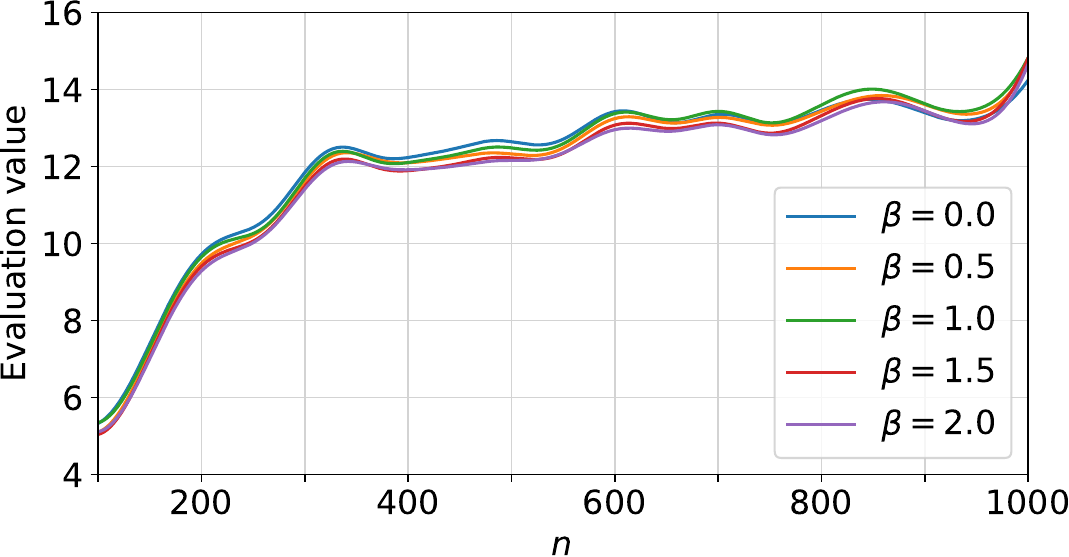} &
    \includegraphics[width=0.4\hsize]{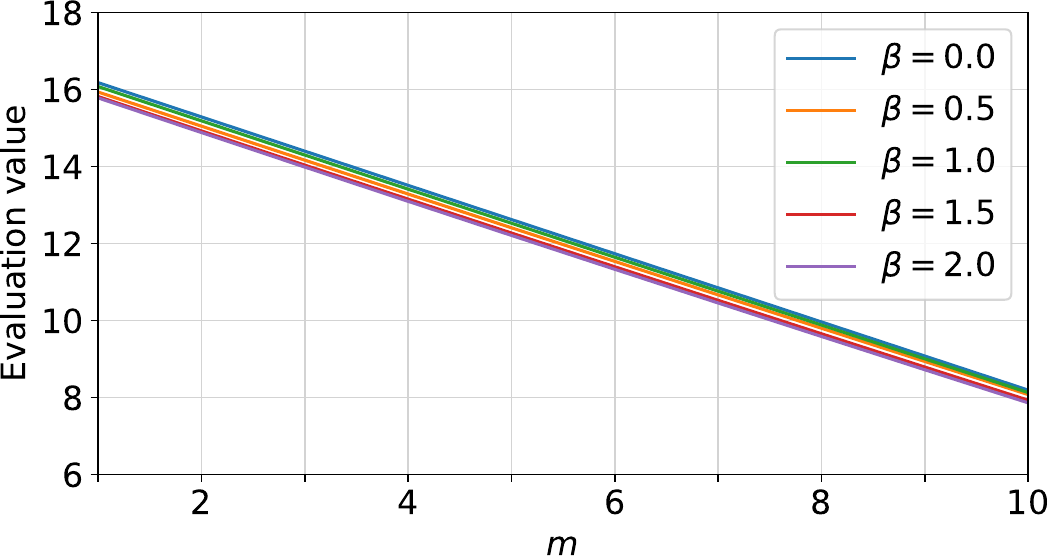} \\
    (a) $n$ & (b) $m$ \\
    \includegraphics[width=0.4\hsize]{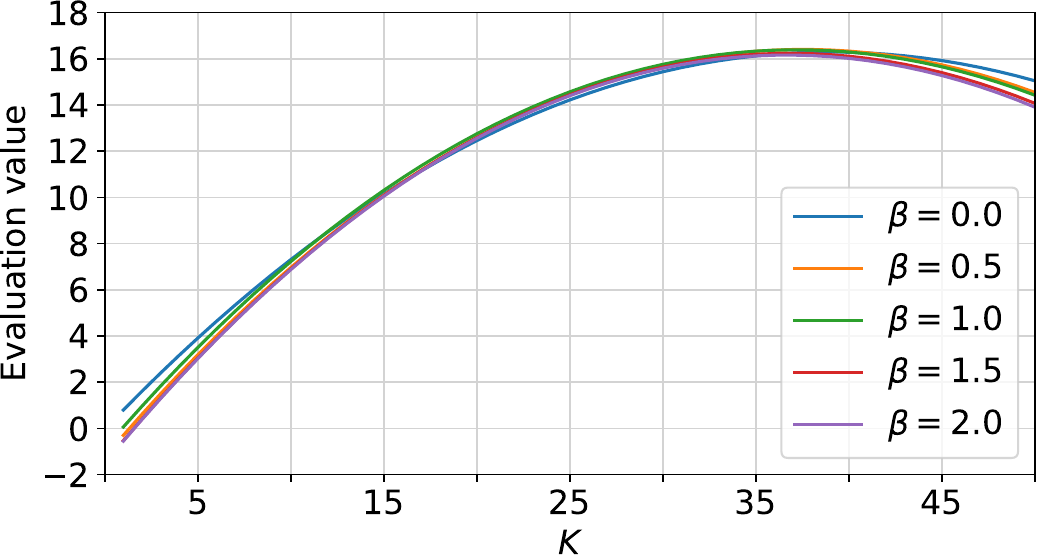} &
    \includegraphics[width=0.4\hsize]{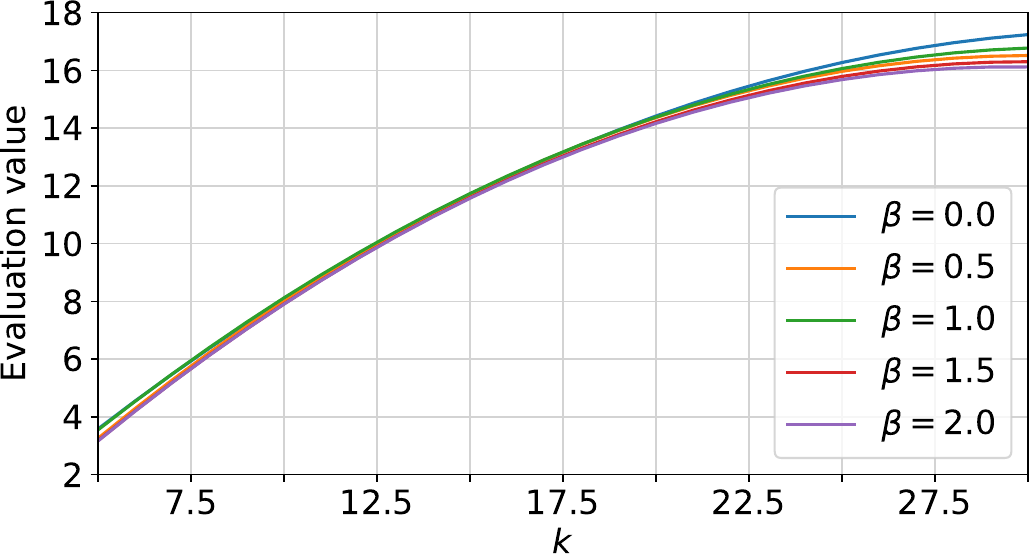} \\
    (c) $K$ & (d) $k$ \\
    \includegraphics[width=0.4\hsize]{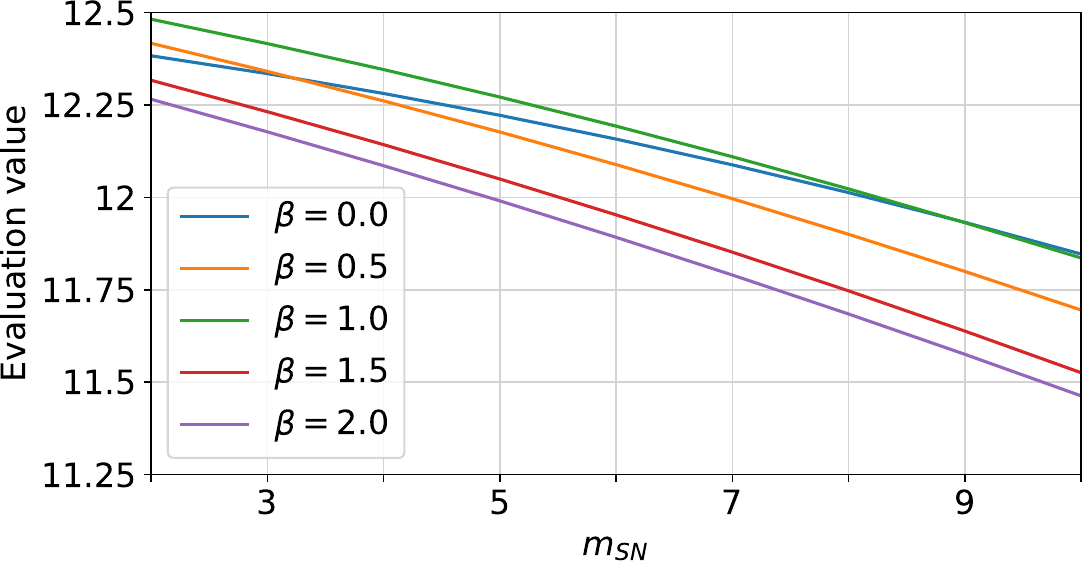} &
    \includegraphics[width=0.4\hsize]{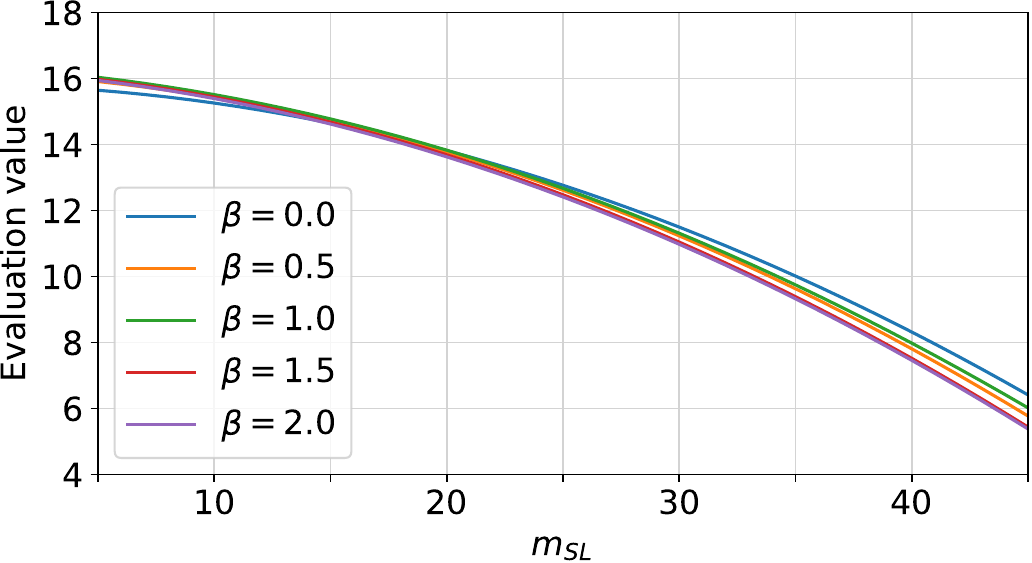} \\
    (e) $m_{SN}$ & (f) $m_{SL}$ \\
    \includegraphics[width=0.4\hsize]{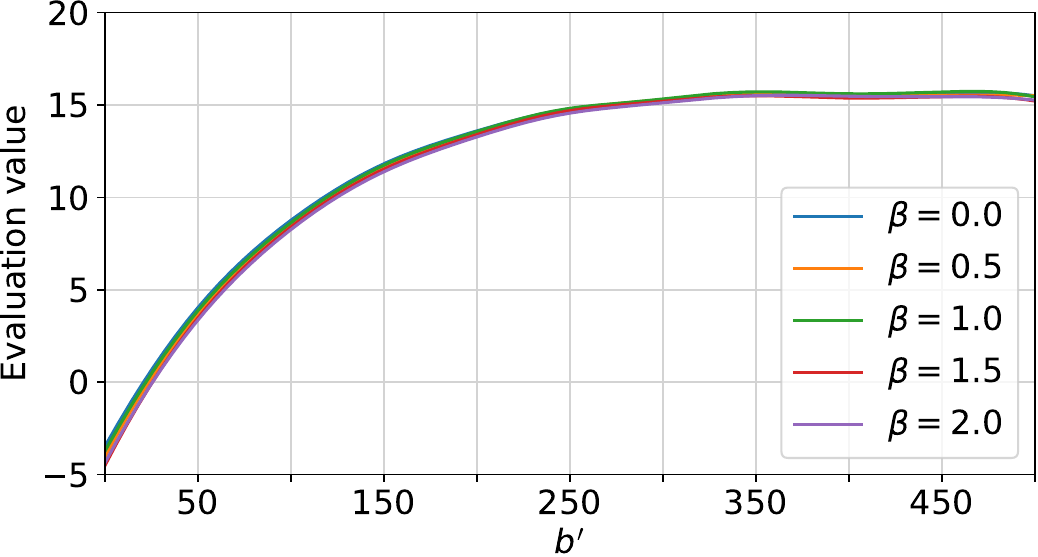} &
    \includegraphics[width=0.4\hsize]{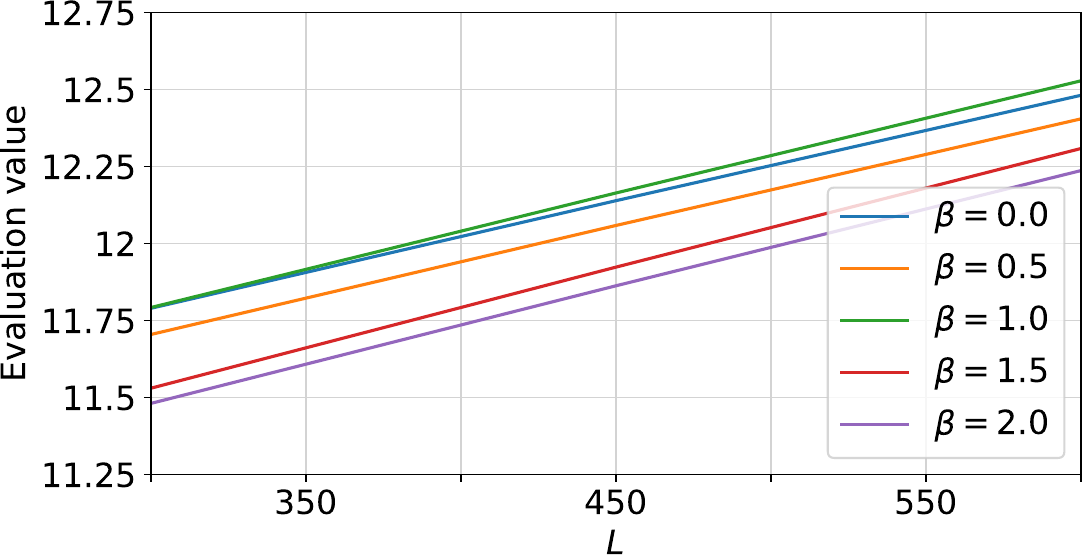} \\
    (g) $b^{\prime}$ & (h) $L$
  \end{tabular}
  \caption{Estimated evaluation values for each experimental parameter at different $\beta$}
  \label{fig:beta}
\end{figure*}

%% file: tex/section6.tex
\section{Conclusion}
In this study, we defined a crowdsourcing team formation problem, termed the multi-team formation problem, and proposed a heuristic team formation algorithm.
In addition to the skill level and budget constraints considered in previous studies, our team formation problem introduced a team size requirement.
The objective function is based on density -- a plausible measure of team compatibility derived from social networks.
Furthermore, the problem assumed concurrent team formation for multiple tasks while limiting the worker load.

The simulation experiment verified both the characteristics and effectiveness of the proposed algorithm.
GPR was used to elucidate the relationship between the eight experimental parameters and the evaluation values.
The results suggest that low values for the team size limit, degree mean, and additional budget adversely affect the evaluation values.
We determined that a large number of workers is necessary when both the number of tasks and the distribution mean of required skill levels are high.
The proposed algorithm has two hyperparameters: $\alpha$, which controls the rate of temperature decrease in the simulated annealing method, and $\beta$, which controls the smoothing degree.
According to this experiment, the highest evaluation values were observed when $\alpha = 0.9$, while variations in $\beta$ had minimal impact.
This suggests that smoothing process implemented in the proposed algorithm is ineffective.
A comparison between the proposed algorithm and the hill-climbing method demonstrated the effectiveness of the annealing method.

One limitation of this study is that we compared the proposed algorithm only with a hill-climbing method as a baseline, a simple heuristic.
While this sufficed to demonstrate the superiority of the proposed  algorithm, more comprehensive comparisons with other metaheuristics such as genetic algorithms or reinforcement learning methods remain for future work.

There are two primary areas for future research in this field.
The first concerns the handling of new workers in the system.
In the proposed algorithm, workers who are closely connected on a social network are more likely to be selected as team members.
Consequently, well-connected workers tend to be chosen more often, whereas those with fewer neighboring nodes are less likely to be selected.
As teamwork is implemented, the edges of the social network are updated; however, a newly registered crowd worker typically has little experience with teamwork and few connections.
This raises the challenge of fairly integrating new workers into the system.
One possible approach is to adjust communication costs for edges connected to new workers, either by slightly increasing them to reflect uncertainty in their compatibility, or by lowering them to give such workers additional opportunities to be selected.
Alternatively, predictive models could be employed to predict and augment their network connections based on worker profiles and the overall social network structure.
Addressing how to balance fairness for new workers with the reliability of team performance remains an important direction for future work.
The second area involves developing a program to generate initial solutions for the proposed algorithm.
Before applying an optimization algorithm based on simulated annealing, it is essential to determine whether feasible solutions that satisfy all constraints exist.
The current approach addresses this issue using an external SAT solver; however, this solver requires significant computation time when the search space is large and the region of feasible solutions is small.
In addition, a large solution space combined with excessive constraints can lead to file encoding errors.
Therefore, we are currently developing a new algorithm to solve this constraint satisfaction problem, as described in \cite{2024_r.yamamoto}.